\input harvmac

\input epsf.tex
\ifx\epsfbox\UnDeFiNeD\message{(NO epsf.tex, FIGURES WILL BE IGNORED)}
\def\figin#1{\vskip2in}
\else\message{(FIGURES WILL BE INCLUDED)}\def\figin#1{#1}\fi
\def\ifig#1#2#3{\xdef#1{fig.~\the\figno}
\midinsert{\centerline{\figin{#3}}%
\smallskip\centerline{\vbox{\baselineskip12pt
\advance\hsize by -1truein\noindent{\bf Fig.~\the\figno:} #2}}
\bigskip}\endinsert\global\advance\figno by1}

\noblackbox
%
%
\font\tenbifull=cmmib10 
\font\tenbimed=cmmib10 scaled 800
\font\tenbismall=cmmib10 scaled 666
\textfont9=\tenbifull \scriptfont9=\tenbimed
\scriptscriptfont9=\tenbismall

\def\IZ{\relax\ifmmode\mathchoice
{\hbox{\cmss Z\kern-.4em Z}}{\hbox{\cmss Z\kern-.4em Z}}
{\lower.9pt\hbox{\cmsss Z\kern-.4em Z}}
{\lower1.2pt\hbox{\cmsss Z\kern-.4em Z}}\else{\cmss Z\kern-.4em Z}\fi}

\font\cmss=cmss10 \font\cmsss=cmss10 at 7pt

\font\ttsmall=cmtt10 at 8pt

\def\yboxit#1#2{\vbox{\hrule height #1 \hbox{\vrule width #1
\vbox{#2}\vrule width #1 }\hrule height #1 }}
\def\fillbox#1{\hbox to #1{\vbox to #1{\vfil}\hfil}}
\def\ybox{{\lower 1.3pt \yboxit{0.4pt}{\fillbox{8pt}}\hskip-0.2pt}}

\def\comments#1{}

\def\CN{{\cal N}}

\def\CL{{\cal L}}

\def\a{\alpha}
\def\ta{\tilde\alpha}

\def\II{\relax{I\kern-.07em I}}
\def\IIA{{\II}A}
\def\IIB{{\II}B}

\def\inbar{\,\vrule height1.5ex width.4pt depth0pt}
\def\IZ{\relax\ifmmode\mathchoice
{\hbox{\cmss Z\kern-.4em Z}}{\hbox{\cmss Z\kern-.4em Z}}
{\lower.9pt\hbox{\cmsss Z\kern-.4em Z}}
{\lower1.2pt\hbox{\cmsss Z\kern-.4em Z}}\else{\cmss Z\kern-.4em
Z}\fi}
\def\IB{\relax{\rm I\kern-.18em B}}
\def\IC{{\relax\hbox{$\inbar\kern-.3em{\rm C}$}}}
\def\ID{\relax{\rm I\kern-.18em D}}
\def\IE{\relax{\rm I\kern-.18em E}}
\def\IF{\relax{\rm I\kern-.18em F}}
\def\IG{\relax\hbox{$\inbar\kern-.3em{\rm G}$}}
\def\IGa{\relax\hbox{${\rm I}\kern-.18em\Gamma$}}
\def\IH{\relax{\rm I\kern-.18em H}}
\def\IK{\relax{\rm I\kern-.18em K}}
\def\IP{\relax{\rm I\kern-.18em P}}

\font\cmss=cmss10 \font\cmsss=cmss10 at 7pt
\def\IR{\relax{\rm I\kern-.18em R}}

\def\fk{{\bf k}}

%

\def\NP{{\it Nucl. Phys.\ }}

\def\PL{{\it Phys. Lett.\ }}
\def\PR{{\it Phys. Rev.\ }}
\def\PRL{{\it Phys. Rev. Lett.\ }}

\def\Mod{{\it Mod. Phys. Lett.\ }}

%
%
\lref\joe{ J. Polchinski, \PRL {\bf 75} (1995) 4724, hep-th/9510017 and
``TASI Lectures on D-branes,'' preprint NSF-ITP-96-145 hep-th/9611050.}
\lref\kkv{ S. Katz, A. Klemm and C. Vafa, ``Geometric 
Engineering of
Quantum Field Theories,'' hep-th/9609239.}
\lref\bunch{ S. Katz and C. Vafa, ``Geometric Engineering of N=1
Quantum Field Theories,''  hep-th/9611090.
M. Bershadsky, A. Johansen, T. Pantev, V. Sadov and C. Vafa,
``F-theory, Geometric Engineering and N=1 Dualities,'' hep-th/9612052.
C. Vafa and B. Zwiebach, ``N=1 Dualities of SO and USp Gauge Theories and
T-Duality of String Theory,'' hep-th/970101.
H. Ooguri and C. Vafa, ``Geometry of N=1 Dualities in Four Dimensions,''
hep-th/9702180.
C. Ahn and K. Oh, ``Geometry, D-Branes and N=1 Duality in Four
Dimensions I,'' hep-th/9704061.
C. Ahn, ``Geometry, D-Branes and N=1 Duality in Four Dimensions II,''
hep-th/9705004.
C. Ahn, R. Tatar, ``Geometry, D-branes and N=1 Duality in Four
Dimensions with Product Gauge Group,'' hep-th/9705106.}
\lref\hw{A. Hanany and E. Witten,``Type IIB Superstrings, BPS
Monopoles and Three-Dimensional Gauge Dynamics,'' hep-th/9611230.}
\lref\deboer{J. de Boer, K. Hori, H. Ooguri and Y. Oz, ``Mirror
Symmetry in Three Dimensional Gauge Theories, SL(2,Z) and D=Brane
Moduli Spaces,'' hep-th/9612131, J. de Boer, K. Hori, Y. Oz and Z. Yin,
``Branes and Mirror Symmetry in $\CN\!=\!2$ Supersymmetric Gauge Theories in
Three Dimensions,'' hep-th/9702154.}
\lref\egk{S. Elitzur, A. Giveon and D. Kutasov, ``Branes and N=1
Duality In String Theory,'' hep-th/9702014.}
\lref\egkrs{S. Elitzur, A. Giveon, D. Kutasov, E. Rabinovici and
A. Schwimmer, ``Brane Dynamics and N=1
Supersymmetric Gauge Theory,''  hep-th/9704104.}
\lref\ejs{ N. Evans, C. V. Johnson and A. Shapere, ``Orientifolds,
Branes and Duality of 4D Gauge Theories,'' hep-th/9703210.}
\lref\bsty{A. Brandhuber, J. Sonnenschein, S. Theisen and S.
Yankielowicz, ``Brane Configurations and 4D Field Theories,'' hep-th/9704044.}
\lref\supered{E. Witten, ``Solutions of four-dimensional Field
Theories via M-Theory,'' hep-th/9703166.}
\lref\andi{A. Strominger, ``Open P-Branes,'' \PL {\bf B383} (1996) 44,
hep-th/9512059; P. Townsend, ``D-Branes from M-Branes,'' \PL {\bf
B373} (1996) 68, hep-th/9512062.}
\lref\se{K. Intriligator and N. Seiberg, ``Mirror Symmetry in Three
Dimensional Gauge Theories,'' \PL {\bf B387}, (1996) 513,
hep-th/9607207.}
\lref\sei{N. Seiberg, ``Electric-Magnetic Duality in Supersymmetric
Non-Abelian Gauge
Theories,'' \NP {\bf435} (1995) 129, hep-th/9411149,
K. Intriligator and N. Seiberg, ``Lectures on supersymmetric gauge
theories and electric-magnetic duality,'' {\it Nucl. Phys. Proc. Suppl.}
 {\bf 45BC} (1996) 1, hep-th/9509066.}
\lref\sw{N. Seiberg and E. Witten, ``Monopole Condensation, And Confinement
In $\CN=2$ Supersymmetric Yang-Mills Theory,'' \NP {\bf 426} (1994) 19 hep-th
9407087 and ``Monopoles, Duality and Chiral Symmetry Breaking in $\CN=2$
Supersymmetric QCD,'' \NP {\bf 431} (1994) 484, hep-th/9408099.}
\lref\klty{A. Klemm, W. Lerche, S. Theisen, and S. Yankielowicz, ``Simple
Singularities and N=2 Supersymmetric Yang-Mills,''
\PL {\bf B344} (1995) 169, hep-th/9411048.}
\lref\af{P. Argyres and A. Faraggi, ``The Vacuum Structure and Spectrum of
N=2 Supersymmetric SU(N) Gauge Theory,'' \PRL {\bf 73} (1995) 3931,
hep-th/9411057.}
\lref\ho{A. Hanany and Y. Oz, ``On the Quantum Moduli Space of Vacua
of $N=2$ Supersymmetric $SU(N_c)$ Gauge Theories,''
\NP {\bf B452} (1995) 283, hep-th/9505075.}
\lref\aps{P. Argyres, M. Plesser, and A. Shapere, ``The Coulomb Phase of
N=2 Supersymmetric QCD,'' \PRL {\bf 75}(1995)
1699, hep-th/9505100.}
\lref\ds{U. Danielsson and B. Sundborg, ``The Moduli Space and
Monodromies of N=2 Supersymmetric SO(2r+1) Yang-Mills Theory,''
\PL {\bf B358} (1995) 273, hep-th
9504102.}
\lref\bl{A. Brandhuber and K. Landsteiner,``On the Monodromies of
N=2 Supersymmetric Yang-Mills Theory
with Gauge Group SO(2n),'' \PL {\bf 358} (1995) 73, hep-th
9507008.}
\lref\ha{A. Hanany, ``On the Quantum Moduli Space of N=2 Supersymmetric
Gauge
Theories,'' \NP {\bf B466} (1996) 85, hep-th/9509176.}
\lref\as{P. C. Argyres and A. D. Shapere, ``The Vacuum Structure of N=2
SuperQCD with Classical Gauge
Groups,'' \NP {\bf B461}(1996) 437, hep-th
9509175.}
\lref\bbs{K. Becker, M. Becker and A. Strominger, \NP {\bf 456} (1995),
130, hep-th 9509175.}
\lref\lwlpg{W. Lerche and N. Warner, `` Exceptional SW Geometry from
ALE Fibrations,'' hep-th/9608183, K. Landsteiner, J. M. Pierre, S. B.
Giddings, ``On the Moduli Space of N = 2 Supersymmetric $G_2$ Gauge Theory,''
\PR {\bf D55} (1997) 2367.}
\lref\klmvw{A. Klemm, W. Lerche, P. Mayr, C. Vafa and N. Warner, ``Self-Dual
Strings and N=2 Supersymmetric Field Theory,''
\NP {\bf B477}
(1996) 746, hep-th/9604034.}
\lref\mw{E. Martinec and N. Warner, ``Integrable systems and supersymmetric
gauge theory,'' \NP {\bf 459} (1996) 97, hep-th
9509161.}
\lref\kdp{E. D'Hoker, I.M. Krichever and D.H. Phong, ``The Effective
Prepotential of N=2 Supersymmetric $SO(N_c)$ and $Sp(N_c)$ Gauge
Theories,'' \NP {\bf B489} (1997) 211,
hep-th/9609145.}
\lref\GP{E. Gimon and J. Polchinski, ``Consistency Conditions for
Orientifolds and D-Manifolds,'' \PR {\bf D54} (1996) 1667, hep-th
9601038.}
\lref\edanom{E. Witten, ``An $SU(2)$ Anomaly,'' \PL {\bf 117B} (1982)
324.}
\lref\klemm{A. Klemm, ``On the Geometry behind N=2 Supersymmetric
Effective
Actions in Four Dimensions,'' hep-th/9705131.}
\lref\cliff{C.V. Johnson, ``On the Orientifolding of Type II
NS-Fivebranes,'' hep-th/9705148.}
\lref\gibbons{G.W Gibbons and P. Rychenkova, ``HyperKahler Quotient
Construction of BPS Monopole Moduli Spaces,'' hep-th/9608085.}%
\lref\hawking{S. Hawking, ``Gravitational Instantons,'' \PL {\bf 60A}
(1977) 81.}
\lref\ptowns{P. Townsend, ``The Eleven-Dimensional Supermembrane
Revisited,'' \PL {\bf B350} (1995) 184, hep-th/9501068.}
\lref\ahha{O. Aharony and A. Hanany, ``Branes, Superpotentials and
Superconformal Fixed Points,'' hep-th/9704170.}
\lref\brha{J. H. Brodie and A. Hanany, ``Type IIA Superstrings,
Chiral Symmetry, and N=1 4D Gauge Theory Dualities,'' hep-th/9704043.}
\lref\donagiwit{R. Donagi and E. Witten, ``Supersymmetric Yang-Mills
and Integrable Systems,'' \NP {\bf B460} (1996) 299, hep-th/9510101.}
\lref\hitchin{N. Hitchin, ``The Self-Duality Equations on a Riemann
  Surface,'' {\it Proc. London Math. Soc.} {\bf 55} (1987) 59,
  ``Stable Bundles and Integrable Systems,'' {\it Duke Math. J.} {\bf
    54} (1987) 91.}
\lref\nakatsu{T. Nakatsu and K. Takasaki, ``Whitham-Toda Hierarchy and
  N=2 Supersymmetric Yang-Mills Theory,'' \Mod {\bf A11} (1996) 157,
  hep-th/9509162.}
\lref\russians{A. Marshakov, ``On Integrable Systems and
  Supersymmetric Gauge Theories,'' hep-th/9702083, and references therein.}
\lref\donagiein{R. Donagi, L. Ein and R. Lazarsfeld, ``A non-linear
  deformation of the Hitchin dynamical system,'' alg-geom/9504017.}
\lref\ta{R. Tartar, ``Dualities in 4D Theories with Product Gauge Groups
from Brane Configurations'', hep-th/9704198.}
%
%
\Title{\vbox{
\hbox{NSF-ITP-97-052}
\hbox{UCSBTH-97-12}
\hbox{CALT-68-2119}
\hbox{\tt hep-th/9705199}
}}{\vbox{\centerline {$\CN\!=\!2$ Supersymmetric Gauge Theories,}
\vskip2pt\centerline{ Branes and Orientifolds} }}
\bigskip
\centerline{Karl Landsteiner$^{a}$ , Esperanza Lopez$^{b}$
and David A. Lowe$^{c}$}\footnote{}{\ttsmall
$^a$ karll@cosmic1.physics.ucsb.edu, $^b$ elopez@itp.ucsb.edu, $^c$
lowe@theory.caltech.edu}\footnote{}{$^c$ Address after August 1, 1997, Physics
Dept., Brown University, Providence, R.I. 02912, USA}

\bigskip\centerline{\it $^{a}$Physics Department, University of California,
Santa Barbara, CA 93107, USA}
\centerline{\it $^{b}$Institute of Theoretical Physics, University of
California, Santa Barbara, CA 93107, USA}
\centerline{\it $^{c}$California Institute of Technology, Pasadena, CA 91125,
USA}

\vskip 1cm
\centerline{\bf Abstract}

Starting with configurations of fourbranes, fivebranes, sixbranes and
orientifolds in Type \IIA\ string theory we derive via M-theory the curves
solving $\CN\!=\!2$ supersymmetric gauge theories with gauge groups $SO(N)$
and $Sp(2N)$. We also obtain new curves describing theories with product
gauge groups. A crucial role in the
discussion is played by the interaction of the orientifolds with the
NS-fivebranes.
\Date{\vbox{\hbox{\sl {May 1997}}}}
\goodbreak

\newsec{Introduction}

Recently it has become increasingly clear that D(irichlet)-branes \joe\ are an
extremely powerful tool for studying supersymmetric gauge theories. In fact
many phenomena are actually better understood from the D-brane point of view.
There are essentially two philosophies for studying gauge theories with
D-branes. One is to compactify string theory on a Calabi-Yau
space. The BPS-states of the gauge theory can then be identified as the
wrapping modes of certain branes around the
non-trivial homology cycles of the Calabi-Yau space \klmvw. A recent review 
of this geometrical engineering \kkv\ of gauge theories can be found in
\klemm.
A different
approach has been pioneered in \hw\ and further developed in \refs{\deboer
\egk \ejs \brha \bsty \egkrs \ahha {--} \ta} and especially in \supered. 
Here one utilizes the
fact that branes themselves can end on branes \andi\ to construct
gauge theories from intersecting branes in a flat background spacetime. More
precisely,
D-threebranes in Type \IIB\ or D-fourbranes in Type \IIA\ string theory can
stretch between a pair of NS fivebranes. Since the the D-branes are finite
in one direction the effective world volume gauge theory will be
three-dimensional in the \IIB\ case or four-dimensional in the \IIA\ case.
Using the $SL(2,Z)$ symmetry of \IIB\ string theory one can then
rederive \refs{\hw, \deboer}
the recently discovered mirror symmetry of three-dimensional
supersymmetric gauge theories \se. On the Type \IIA\ side
the dualities \sei\ which are
characteristic for $\CN=1$ supersymmetric gauge theories in four
dimensions\foot{For a
geometrical engineering approach to $\CN=1$ theories see \bunch.} have been
recovered \refs{\egk \ejs \brha \bsty \egkrs \ahha {--} \ta}.

$\CN\!=\!2$ supersymmetric gauge theories in four dimensions were the
subject of interest in
\supered. The Coulomb branch of the moduli space of four-dimensional
$\CN\!=\!2$ gauge theories coincides with the moduli space of a
particular family of Riemann surfaces. This was first shown in the
ground-breaking work of Seiberg and Witten \sw\ for the case of the
gauge group $SU(2)$. In a series of papers this has been
generalized to the other classical gauge groups
\refs{\klty\af\ho\ds\aps\bl\ha{--}\as}\foot{The treatment of exceptional
gauge groups has posed more
problems \lwlpg.}. The construction of Witten in \supered\ lifts a
configuration of fourbranes stretched between fivebranes in Type \IIA\
string theory to M-theory. The important feature here is that the
fourbranes themselves are secretly fivebranes of M-theory wrapped
around the eleventh dimension. The configuration of intersecting branes
in ten dimensions is a projection of an eleven-dimensional
configuration with a single fivebrane wrapped around a two-dimensional
Riemann surface. The effective worldvolume theory on the fivebrane
becomes four dimensional and is precisely the $\CN\!=\!2$ gauge theory
whose low-energy effective action is described by that same Riemann
surface. In the context of gauge theories
these Riemann surfaces have been derived by a good-guess ansatz which
could be justified by performing some highly nontrivial, physical
consistency tests. The power of the brane construction lies in the fact
that it gives a physical and comparatively straightforward derivation
from ``first principles'' of
these Riemann surfaces!  While Witten restricted himself to the case of
unitary gauge groups and products thereof this paper is devoted to the
generalization these results to the case of the classical gauge groups
$SO(N)$ and $Sp(2N)$. The new ingredient that we need
is an orientifold. In adding an orientifold we follow recent
work in \ejs. $\CN\!=\!1$ dualities from this point of view have been
discussed in \refs{\ejs, \egkrs}. The role played by the Riemann surfaces in
describing the nonperturbative corrections to the classical brane
configurations in situations with $\CN\!=\!2$ has been pointed out
also in \ejs.
Our aims differ in that we are solely interested in
deriving the Riemann surfaces of the $\CN\!=\!2$ theories from D-brane
considerations. In the course
of doing this, using information from Type \IIA\ string theory,
M-theory and $\CN\!=\!2$ gauge theories, we will obtain a
rather interesting picture of the interaction of NS-fivebranes with an
orientifold fourplane.

We will review Witten's construction \supered\ in
section two. Section three deals
with the addition of an orientifold fourplane. Although an orientifold
is an intrinsically stringy object we will argue that many of its
properties (at least the properties crucial to our construction) carry
over to M-theory. We derive the Riemann surfaces for the $\CN\!=\!2$ theories
with gauge groups $SO(N)$ and $Sp(2N)$ with and without matter
hypermultiplets transforming in the fundamental representation of the
gauge group. Witten pointed out that there are two mechanisms (albeit
connected by a phase transition) for adding hypermultiplets. One
consists of adding fourbranes stretching from a NS-fivebrane off to
infinity, the other by adding D-sixbranes. The latter will be the subject
of section four. Section five deals with generalizations of the
elliptic models of \supered. Many new theories with vanishing
beta-function emerge.

\newsec{Review of Witten's construction}

\ifig\basicfig{A configuration of fivebranes connected by parallel
fourbranes.}
{
\epsfxsize=3.2truein\epsfysize=3.0truein
\epsfbox{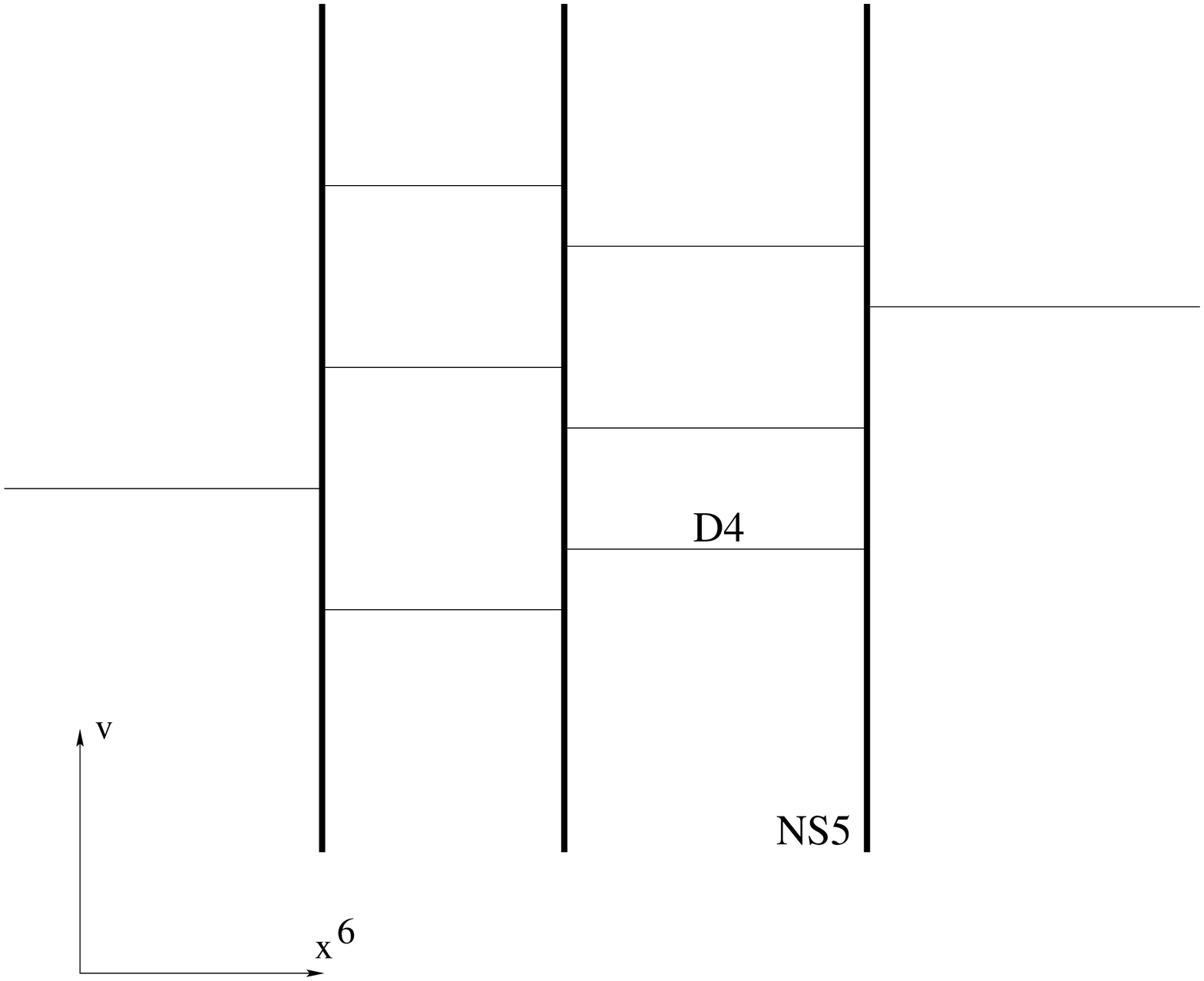}}

The basic configuration of intersecting branes is schematically
depicted in
\basicfig.
The fivebranes\foot{We follow in our nomenclature \supered\ and speak
only of four- five- and sixbranes dropping the specification NS or D
since in Type \IIA\ the dimension of the brane determines if it is NS
or D.} extend in the directions $x^0,x^1,\cdots,x^5$, are located at
$x^7 = x^8 = x^9 =0$ and at some arbitrary value of $x^6$. The latter
is only well defined in the classical approximation. We introduce the
complex variable $v = x^4 + i x^5$. The fourbranes are stretched
between the fivebranes. They extend over
$x^0,\cdots,x^3$,
live (classically) at a point in the $v$-plane and are finitely
extended
in $x^6$. Since they are stretched between the fivebranes they live at
the
same point in the remaining dimensions.
Sixbranes may also be present. These extend then in the
directions $x^0,\cdots,x^3,x^7,x^8,x^9$ and live at a point in $x^4,x^5,x^6$.

The end of a fourbrane looks like a vortex on the fivebrane worldvolume. The
position of the fivebrane in the $x^6$ direction becomes
a scalar field in the worldvolume theory and obeys
\eqn\poisson{
\nabla^2 x^6 = \sum_i q_i \delta^2 (v-a_i) .}
Here the $a_i$ are the positions of the fourbranes ending on the fivebrane in
consideration. The charges $q_i$ are either $+1$ or $-1$ depending if the
corresponding fourbrane extends to the left or to the right of the fivebrane in
the $x^6$ direction. In section three we will argue that an orientifold
fourplane parallel to the fourbranes induce charges $\pm 2$.

Because the fourbrane extends over three dimensions of the fivebrane, equation
\poisson\ is effectively a two-dimensional Poisson equation with solution
\eqn\solx{
x^6 = k \sum_i q_i \log | v-a_i | . }
These formulas are only valid in a kind of semiclassical approximation where
one is at large $|v|$ and simultaneously far away from each fourbrane. A
fivebrane has a well defined $x^6$-value for $v\rightarrow \infty$ only if it
is neutral, i.e. $\sum q_i =0$. The constant $k$ depends only on the Type
\IIA\ string coupling constant.

The contribution of the ends of the fourbrane to the energy of a
fivebrane is given by a term in the action of the fivebrane of the
form $\int\,d^4x\,d^2v \partial_\mu x^6 \partial^\mu x^6$,
where the indices $\mu$ run over $0,\cdots,3$. For this to be finite it
is necessary that
\eqn\finite{
\sum_i a_i - \sum_j b_j = {\rm const.},}
where $a_i$ are the positions of the fourbranes to the left and $b_j$ the
positions of the fourbranes to the right.

This configuration is lifted to M-theory by taking into account the eleventh
dimension, which we denote by $x^{10}$.  It parameterizes a circle of radius
$R$. Introducing the complex coordinate $s = (x^6 + i x^{10})/R$ equation
\solx\ is generalized to
\eqn\sols{ s = \sum_i q_i \log ( v - a_i) .}

If there are $k_\a$ fourbranes in between the $\a$-th and ($\a\!+\!1$)-th
fivebranes we will get a $SU(k_\a)$ gauge theory in the four dimensions
$x^0,\cdots,x^3$. The overall $U(1)$ factor, usually present in gauge theories
realized by D-branes, is frozen out. This is essentially due to the finite
energy condition \finite. Fourbranes to the left of the $\a$-th fivebrane and
to the right of the ($\a\!+\!1$)-th fivebrane give rise to hypermultiplets in
the fundamental of $SU(k_\a)$. In a configuration with $n+1$ fivebranes the
gauge group is thus $\prod_{\a=1}^n SU(k_\a)$ with matter content transforming
in the
$(\fk_1,\bar{\fk}_2)\oplus(\fk_2,\bar{\fk}_3)
\oplus\cdots\oplus(\fk_{n-1},\bar{\fk}_n)$ (we are assuming
that there are no semi-infinite fourbranes at the ends of
the chain of fivebranes). The constant in the finite
energy condition \finite\ is a characteristic parameter of the $\a$-th
fivebrane giving rise to a bare mass to the hypermultiplets in the
$(\fk_\a,\bar{\fk}_{\a+1})$.
One can also compactify the $x^6$ direction by periodic identification of the
$n\!+\!1$-th fivebrane with the first one. Then there is an overall
$U(1)$-factor in the gauge group.

When we are going to place an orientifold
fourplane we will always work on the covering space. For each fourbrane there
is then a mirror image. This will have two related effects. The constant on
the right hand side of \finite\ is forced to vanish thus there are no bare mass
parameters in the theories with orientifolds. This restriction also
arises from the gauge theory point of view as we see later.
Secondly, even upon compactifying
the $x^6$-direction we do not get an overall $U(1)$ factor in the gauge group.
Although at first glance it might seem so, there is no loss of generality.

The gauge coupling constant of a $SU(k_\a)$ factor is essentially given by the
distance between the $\a$-th and ($\a\!+\!1$)-th fivebranes. Taking into
account the eleventh dimension of M-theory this gives rise to the formula
\eqn\gaugecoupling{- i\tau_\a = s_\a - s_{\a-1} .}
We have $\tau_\a = {\theta_\a\over{2\pi}} + {{4\pi i}\over{g^2_\a}}$. The
difference in the positions of the fivebranes in $x^{10}$ determines the theta
angle of the $\a$-factor of the gauge group. The scale of the gauge theory is
set by $v$. At large $v$ and using $\-i \tau = b_0 \log v$, we can read off the
one loop beta function coefficient $b_{0,\a}$ for the $SU(k_\a)$ factor
\eqn\betaoneloop{b_{0,\a} = -2 k_\a + k_{\a+1} + k_{\a-1} .}

In M-theory the fivebranes and fourbranes are really the same object. What
appears as fourbranes in Type \IIA\ string theory are just M-theory fivebranes
wrapped around the eleventh dimension. Thus the fourbranes are actually better
thought of as tubes connecting the fivebranes. In this way one sees very
directly that there is really only one fivebrane wrapped around a Riemann
surface $\Sigma$. It is embedded in the two-dimensional complex space
parameterized by $v$ and $s$. There is a slight subtlety though: the Riemann
surface obtained in this way extends to some points at infinity and is
therefore noncompact. A compact one can be made by adding a finite number of
points. It has already been shown in \klmvw\ that a fivebrane wrapped around a
Riemann surface gives rise to a $\CN\!=\!2$ $SU(k)$ gauge theory on
the
four-dimensional part of its world volume.
Precisely the same mechanism is present
here. BPS states are M-theory membranes of minimum area whose boundaries live
on $\Sigma$.

The properties of low-energy supersymmetric effective field theory
imply that the low-energy dynamics is described by an
integrable system \donagiwit. Examples of these integrable systems
have been constructed in \refs{\donagiwit\mw \nakatsu{--}\russians}.
The integrable systems that appear in this context
can be described as generalizations of the Hitchin system \hitchin,
which associates an integrable system to a complex curve $\Sigma$
embedded in some complex two-dimensional symplectic manifold $Q$ in
the following way. Let $\Sigma'$ be a curve in $Q$ corresponding to a
deformation of $\Sigma$ and let $\CL'$ be a line bundle on $\Sigma'$.
Then the deformation space of the pairs $(\Sigma', \CL')$ defines an
integrable system \donagiein. They actually considered the
case of compact $\Sigma$, but the results can be extended to the
noncompact case. The gauge theories we consider in this paper
are all described by integrable systems of this type with $Q$
identified appropriately either as flat space $\IC^2$, multi-Taub-NUT
space, or the product $E\times \IC$ ($E$ an elliptic curve), and $\Sigma$
identified with the curve describing the theory.

To compute the genus of the compactified Riemann surface we just have to count
the number of tubes connecting the fivebranes, $g=\sum_\a (n_\a-1)$. Here we
are slightly more general than \supered\ in that we do not necessarily identify
the number of fourbranes $k_\a$ with the number of tubes $n_\a$ connecting the
fivebranes. What we have in mind is the following. On the fivebrane we have an
antisymmetric tensor field with self-dual field strength $T$. Harmonic one
forms $\Lambda$ on $\Sigma$ give rise to gauge fields in four dimensions
through $T =
F\wedge\Lambda + *F\wedge *\Lambda$.
The gauge field is obtained by integrating $T$ over one cycles of $\Sigma$. The
number of harmonic one-forms on a Riemann surface equals its genus and their
period integrals are points on the Jacobian of $\Sigma$. However, as emphasized
in \mw\ the physics of the $\CN\!=\!2$ gauge theory is not determined simply by
the Jacobian of $\Sigma$ but by a sub-abelian variety whose rank coincides with
the rank of the gauge group. This is the so-called special Prym variety
Prym($\Sigma$).  Typically it is the subspace invariant under an involution
of the Riemann surface. Therefore we have to take into account the possibility
that there are more tubes connecting the fivebranes than arise simply
from the fourbranes. Such tubes would be generated by
nonperturbative effects. The
gauge fields originating from them however are expected to play no physical
role and should vanish upon projecting onto Prym($\Sigma$). In our cases this
will be naturally achieved by orientifolding the configuration.

The physically relevant quantity of the low-energy gauge theory that
the curves determine is the matrix of couplings $\tau^{\mu\nu}(u_l)$
as a function of the parameters of the Coulomb moduli space $u_l$, which
appears as the period matrix of the curve. We
take $\mu,\nu = 1,\cdots g$ where $g$ is the genus of the Riemann
surface $\Sigma$.
If one also knows the Seiberg-Witten differential $\lambda$ of the
curve one may also derive the masses of all BPS states
via the equation $M=|a_\mu q^\mu+ a_D^\mu h_\mu|$, where $q^\mu$ are electric
and $h_\mu$ are magnetic quantum numbers and
$a_D$ and $a$ are given by integrals
of $\lambda$, $a_D^\mu =
\oint_{\alpha^\mu} \lambda$, and $a_\nu = \oint_{\beta_\nu} \lambda$.
Here we use a basis of the $2g$ one-cycles $(\a^\mu , \beta_\nu)$
on $\Sigma$ with a
standard intersection form $\langle \a^\mu , \beta_\nu\rangle =
\delta^\mu_\nu$,
$\langle \a^\mu , \a^\nu \rangle = \langle \beta_\mu , \beta_\nu
\rangle =0$.

Following \klmvw\ we may determine the Seiberg-Witten differential for
all the cases considered in this paper.
The BPS states in the fivebrane worldvolume theory arise as membranes
whose boundary lies on the Riemann surface \andi.
The BPS condition requires the worldvolume to be of the form
$\IR\times D$ where $D$ is a complex Riemann surface of minimal area.
Let us consider here the case when the Riemann surface $\Sigma$ is
embedded in flat $Q=\IR^3 \times S^1$. It is convenient to
introduce the single valued variable $t = \exp (-s)$.
The area of $D$ is given by
\eqn\susyarea{
V \sim \biggl| \int_D  {dv ~dt \over t} \biggr| = \biggl|
\int_{\partial D}  {v ~dt \over t}\biggr|~.
}
We should therefore set $\lambda = v ~dt/t$. This differential agrees
with that obtained for $SU(N)$ gauge group with fundamental matter
\refs{\sw\klty\af{--}\ho}. This form for the differential will also
carry over to the all theories considered in the following section.

The precise form of the Riemann surface can be obtained by relatively simple
considerations. It will be described by an equation $F(t,v) = 0$.
For fixed $v$ the zeros of $F(t,v)$ are the positions of the fivebranes.
For fixed $t$ the solutions in $v$ indicate the presence of the
fourbranes. $F(t,v)$ can be determined uniquely by using the information
about the bending of the fivebranes. At large $v$ the solutions in $t$ should
have the form $t_\a = c_\a v^{a_\a}$ where $-a_\a$ is the sum
of the charges
sitting on the $\a$-th fivebrane, $-a_\a = \sum_i q_{i,\a}$ and $c_\a$ is some
constant. Semi-infinite fourbranes sitting to the left or to the right of a
fivebrane will bend this five brane to $x^6 = \pm \infty$. If $v$ equals the
position of such a semi-infinite fourbrane $F(t,v)$ must have $t=0$ or
$t=\infty$ as solution at this particular value of $v$. For a situation with
only two fivebranes and $k$ fourbranes in between them these considerations
determine $F = t^2 + B(v) t + 1$. $B(v)$ is of the form $v^k +
u_2 v^{k-2} +
\cdots u_k$ where a shift in $v$ has been performed to absorb the term
proportional
to $v^{k-1}$. This is equivalent to the curves of \refs{\klty, \af}.
In the following
we will show how one can determine the curves for the orthogonal and symplectic
gauge groups by taking properly into account the effects of the orientifold.

\newsec{Including an Orientifold}

\subsec{The Orientifold}

In this section we want to extend the previous construction to
$\CN\!=\!2$ four-dimensional gauge theories with orthogonal and
symplectic gauge groups.
Orthogonal and symplectic groups can be obtained from D-brane
configurations by introducing an orientifold projection.
This consists of a projection that combines the process of modding out a
space-time symmetry and world-sheet parity inversion. The fixed points of the
space-time symmetry define orientifold planes.

\ifig\orientifig{A configuration of fourbranes and fivebranes and an
orientifold fourplane.}
{
\epsfxsize=3.2truein\epsfysize=3.0truein\epsfbox{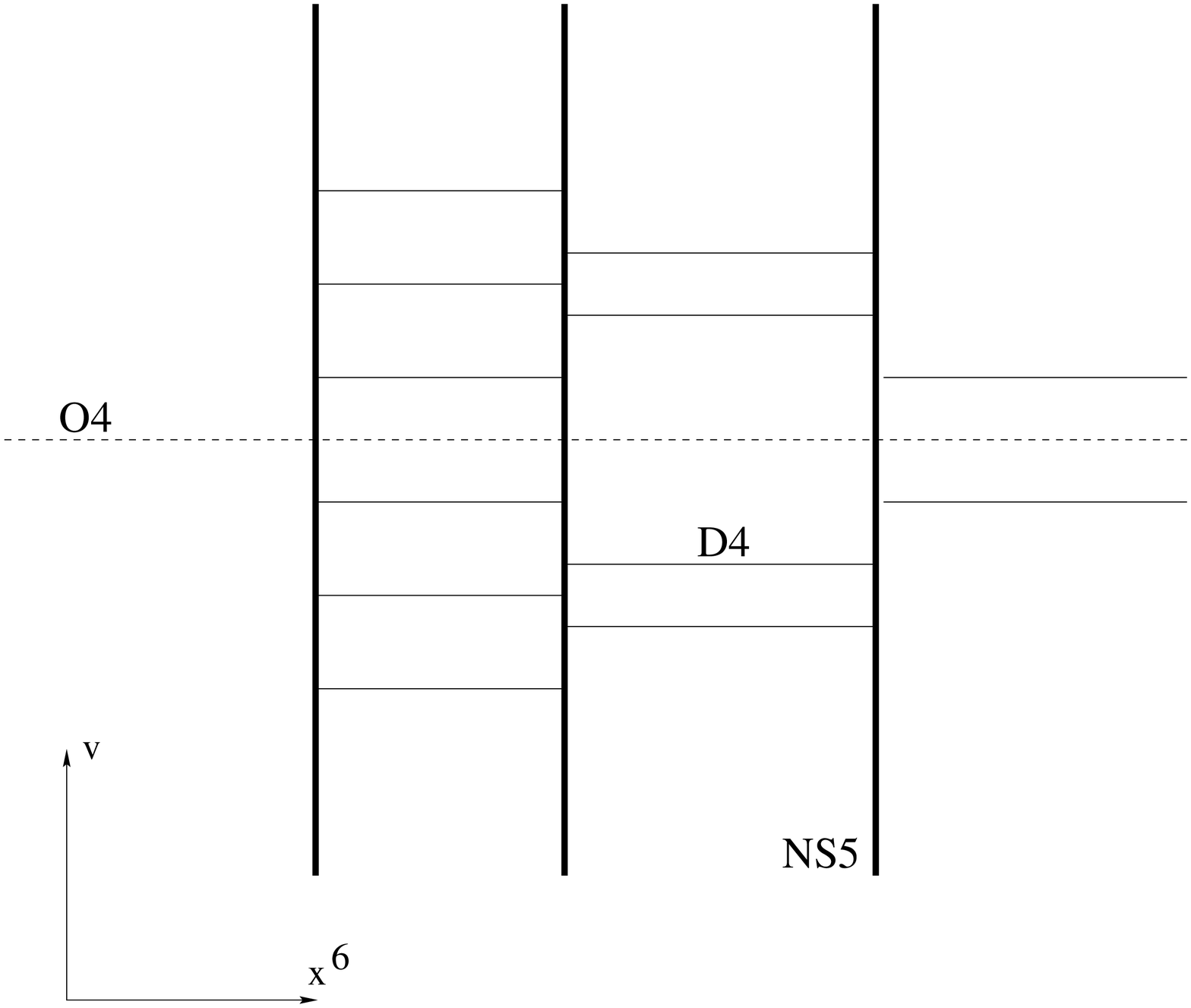}}

Orientifold projections generically break half of the existing
supersymmetries. We can however introduce an orientifold plane
in the D-brane configuration considered in the previous section
without breaking any further supersymmetry. This can be achieved
by placing an orientifold fourplane parallel to the fourbranes\foot{It is
also possible to place an orientifold sixplane parallel to the
sixbranes \refs{\ejs, \egkrs}. In the following however we will not
analyze that case, and will restrict to the presence of an
orientifold fourplane.}.
This corresponds to modding out by the space-time transformation:
\eqn\eoo{(x^4,x^5,x^7,x^8,x^9) \rightarrow
(-x^4,-x^5,-x^7,-x^8,-x^9).}
Each object which does not lie at the fixed point of \eoo,
over the orientifold plane, must have a mirror image.
Since fourbranes joining mirror pairs of fivebranes would break supersymmetry,
in our configurations all fivebranes will sit at
$x^7=x^8=x^9=0$.
Therefore as in the previous section, we can neglect the
$x^7,x^8,x^9$ directions in describing the orientifolded
brane configuration \orientifig.

The world-sheet parity projection $\Omega$ allows for $\Omega^2=\pm 1$.
The sign determines if we will obtain an orthogonal ($\Omega^2=1$) or
symplectic ($\Omega^2=-1$) gauge group \GP.
Orientifold planes behave as non-dynamical BPS objects, carrying
a net charge under RR gauge fields.
The orientifold charge\foot{In a toroidal space $T^5$ \eoo\
would generate 32 orientifold
planes. By sending the radius of each circle to infinity only
one orientifold plane remains at a finite distance. This is
the situation we will consider in the following.}
depends on the type of parity projection,
being $\mp 1$ for $\Omega^2=\pm1$ in the normalization in which
Dirichlet branes have charge $1$.

As argued in \ejs, we will assume that the orientifold extends
along the entire $x^6$-direction and each time it crosses a
fivebrane its charge changes sign. This can be justified by
considering a configuration with $k>2$ fivebranes in which an
orientifold projection has been performed. Fivebranes in the presence
of orientifolds have further been considered in \cliff.
Fourbranes ending to the left of the $\a$-th fivebrane and to the
right of the $\a\!+\!1$-th fivebrane provide fundamental matter for the
gauge theory on the fourbranes in between the $\a$-th and $(\a\!+\!1)$-th
fivebranes. The flavor group for $SO(N_c)$ or $Sp(N_c)$ gauge
theories is constrained to be respectively $Sp(2N_f)$ or $SO(2N_f)$ \as.
Since gauge and flavor groups are interchanged when we move
from one set of fourbranes to the next, we see that the
orientifold should change nature each time it crosses a fivebrane
and should therefore extend along the whole configuration.

In the next subsections we will propose a way of deriving the exact
solution for $\CN\!=\!2$ four-dimensional gauge theories based on
orthogonal and symplectic groups by lifting Type \IIA\ brane
configurations to M-theory along the lines of \supered.
The first question that is raised is how to describe an
orientifold plane in M-theory. Our orientifold fourplane is
charged under the same RR gauge field as Dirichlet fourbranes.
We will assume that, as Dirichlet fourbranes, it corresponds
to a six-dimensional object in M-theory which is wrapped around
the eleventh direction.
The details of this picture will be explained
further in the following.

\subsec{$SO(2k)$ Gauge Groups}

In this subsection we consider orthogonal groups with an even
number of colors, $SO(2k)$. We will restrict to pure Yang-Mills
theories. Our space-time is now an orbifold, we
will however work on the covering space by
considering $\IZ_2$ invariant configurations.

The simplest configuration consists of 2 fivebranes traversed
by the orientifold plane, together with $2k$ fourbranes
ending on the fivebranes.
In order to describe the embedding of this configuration in M-theory
it is sufficient to consider
the 2-complex dimensional space $Q=\{v=x^4+ix^5,
s=x^6+ix^{10}\}$. The orientifold sits at $v=0$, and each fourbrane at
$v$ has a mirror image at $-v$.

We expect that in M-theory, the set
of fourbranes and fivebranes can be described in terms of a single M-fivebrane
wrapped around a certain Riemann surface. Using the variable
$t=e^{-s}$, the Riemann surface will be given in our case by
$F(t,v^2)=0$. Since we have 2 fivebranes, $F$ must be a quadratic
polynomial in $t$
\eqn\eCE{A(v^2) ~t^2 + B(v^2) ~t + C(v^2)=0.}
The first condition this curve must satisfy is to reproduce the
correct bending of each of the fivebranes at large values of $v$. This
bending is determined by the RR charge of the objects that end or
traverse the fivebrane.
The essential ingredient for the determination of $F$ is that the
orientifold plane, though it traverses the fivebranes, provides a
net charge contribution to them. This is due to the fact that the
orientifold charge changes sign when crossing a fivebrane.
In the case we are considering, the orientifold is seen by the
left fivebrane as a $+2$ charge, and as a $-2$ charge \foot{This has also
been discussed in \egkrs.} by the right fivebrane.
Therefore for $v \rightarrow \infty$ \eCE\ should reduce to
\eqn\easOE{t_i \sim v^{a_i} ~ , \quad a_1=-a_2=2k-2,}
where $2k$ is the number of fourbranes ending on each fivebrane.

We analyze now the locus $v=0$, where the orientifold sits.
A semi-infinite fourbrane ending to the left of the
first fivebrane would be represented by \eCE\ as the solution
$t=\infty$ at the value of $v$ where fourbrane and fivebrane, in the
perturbative picture, meet. In the M-theory picture, the whole
configuration is described as a single M-fivebrane. We could then
view the fourbrane as the deformation of the fivebrane induced by
the presence of a non-zero charge in its world-volume.
Following this reasoning, we can treat the interaction between
fivebrane and orientifold in the same way.
Since the orientifold is seen by the fivebrane as a net charge
contribution, it should deform the fivebrane in the direction
determined by the sign of that charge. In our case, this implies
that at $v=0$ the first fivebrane will deform to $t=\infty$ and the
second to $t=0$. Therefore the quadratic equation \eCE\
must have as a solution $v=0$, with $t=0$ and $t=\infty$.
The curve that meets this requirement, together with
\easOE, is
\eqn\eCOE{v^2 ~t^2 + B(v^2) ~t + v^2 =0,}
with $B$ the most general polynomial of order $k$ in $v^2$
\eqn\eB{B(v^2)= v^{2k} + u_2 v^{2k-2} + \cdots + u_{2k}.}
The normalizations in \eCOE\ and \eB\ are fixed by conveniently
rescaling $v$ and $t$. Multiplying \eCOE\ by $v^2$ and redefining
$\tilde{t}=v^2 t + B/2$, the previous curve reads
\eqn\eOE{{\tilde t}^2 = {B(v^2)^2 \over 4} - v^4,}
which is the standard form for the curve that solves $\CN\!=\!2$
$SO(2k)$ gauge theory without matter \bl.
\ifig\orthogfig{Behavior of fivebranes near an orientifold plane
that gives rise to $SO(2k)$ on fourbranes.}
{
\epsfysize=3.0truein
\epsfxsize=3.2truein
\epsfbox{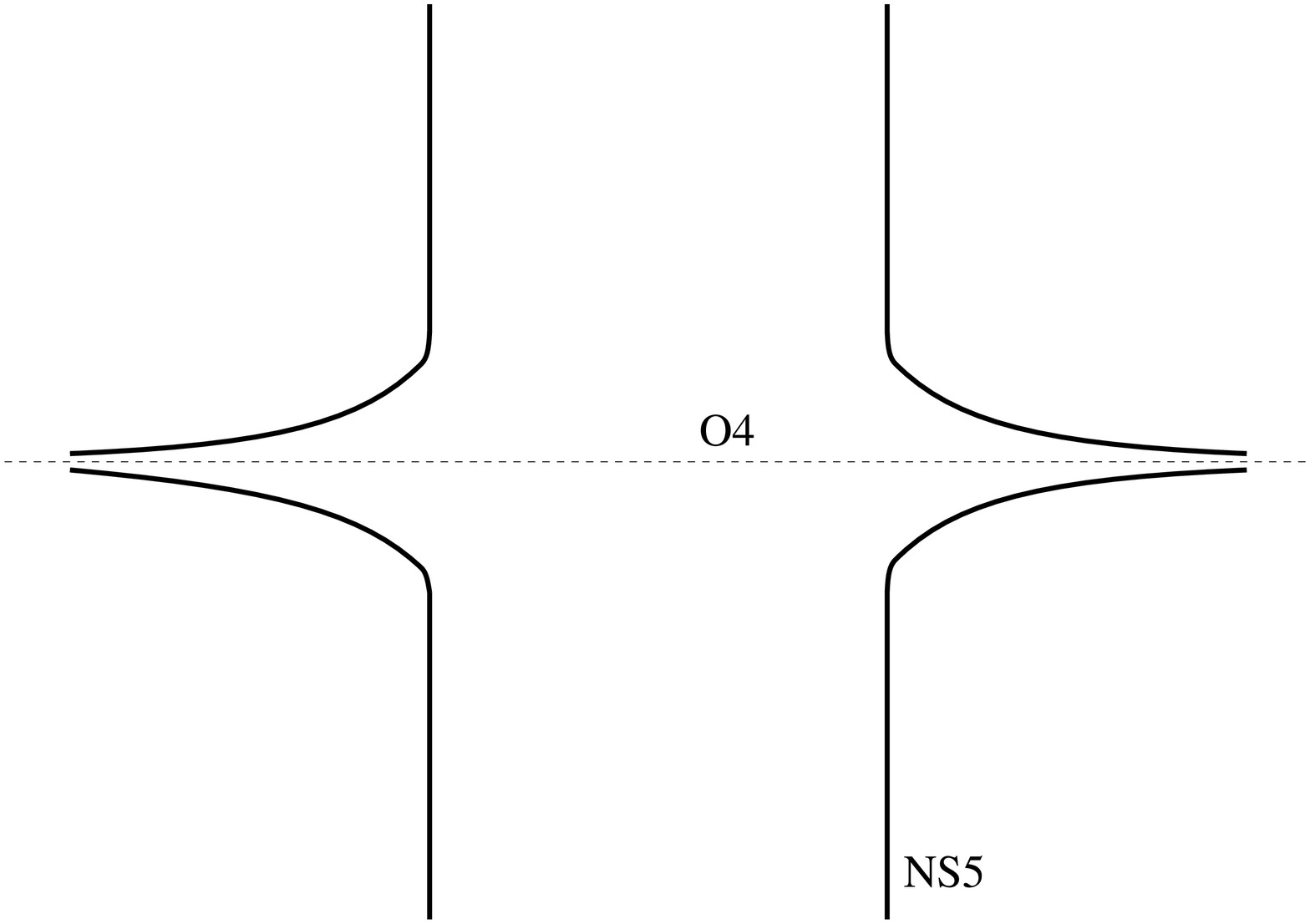}}
%

A somewhat strange feature of this
family of curves is that there are singularities in its moduli space that do
not correspond to massless BPS states. In fact the associated monodromy is
trivial. This singularity sits at $u_{2k} = 0$\foot{ There is a subtlety here
in that the highest Casimir of $SO(2k)$ is actually reducible. More precisely,
due to the structure of the Weyl group $u_{2k} = \tilde{u}^2$. The gauge
invariant quantity $\tilde{u}$ is the trace of the Pfaffian of the
Higgs field
in the adjoint in gauge theory language.}. In the brane configuration this
corresponds to the situation when a mirror pair of fourbranes falls into the
orientifold. The charges of this mirror pair are just enough to cancel the
effective charge of the orientifold on the fivebrane so that the
fivebranes are no longer bent.
The effective gauge coupling squared is proportional to the
inverse distance between the fivebranes. Thus we see that the gauge
coupling at this particular singularity of the Riemann surface stays
finite. This indicates that no additional BPS states become
massless at this point\foot{A finite coupling constant can
also correspond to a singular
point in moduli space with massless vectors and hypermultiplets
whose contributions to the logarithmic divergence cancel each other.
However the case we are analyzing can not provide this matter content,
since the configuration we have does not include semi-infinite
fourbranes or sixbranes.}.
Thus we get a nice physical picture why the singularity of the curve
is not associated with massless states.

\subsec{Symplectic Gauge Groups}

We leave the study of $SO(2k+1)$ for the next subsection, and
consider now symplectic gauge groups, $Sp(2k)$. The reason for
this is that the treatment of $SO(2k)$ and $Sp(2k)$ groups share
many common features.

In particular, we will consider the same configuration of branes
that we used in the previous subsection. Namely, 2 fivebranes on
which $2k$ fourbranes end. Each fourbrane at $v$ will have a mirror
image at $-v$ since we are working on the double cover of an
orbifolded space. The sole difference will be that the
orientifold plane sitting at $v=0$ will have opposite charge
assignments with respect to the $SO(2k)$ case.

Let us analyze first the bending of the fivebranes at large
values of $v$. The orientifold contributes now with a charge
$-2$ to the first fivebrane and a charge $+2$ to the second.
Therefore we have
\eqn\easS{t_i \sim v^{a_i} \;\; , ~~~~~~~~a_1=-a_2=2k+2.}
We notice here a first difference. For unitary and orthogonal
groups $a_2-a_1=b_0$, where $b_0$ was the one-loop
beta function coefficient. However for symplectic groups
we obtain
\eqn\easS{a_2-a_1=-(4k+4)=2 b_0^{Sp}.}

The extra factor $2$ can be explained as a normalization effect
intrinsic to the way in which orthogonal or symplectic groups
are derived from orientifold constructions.

\ifig\cyclefig{(a) The vanishing cycles for the $SU(n)$ case.
(b) After performing an orientifold projection only certain linear combinations
survive. Only two of these cycles are shown. Solid lines correspond to cycles
surviving both $SO(2k)$ and $Sp(2k)$ projections. The dashed line corresponds
to a long root generator
 of $Sp(2k)$.}
{
\epsfxsize=3.5in
\epsfysize=1.8in
\epsfbox{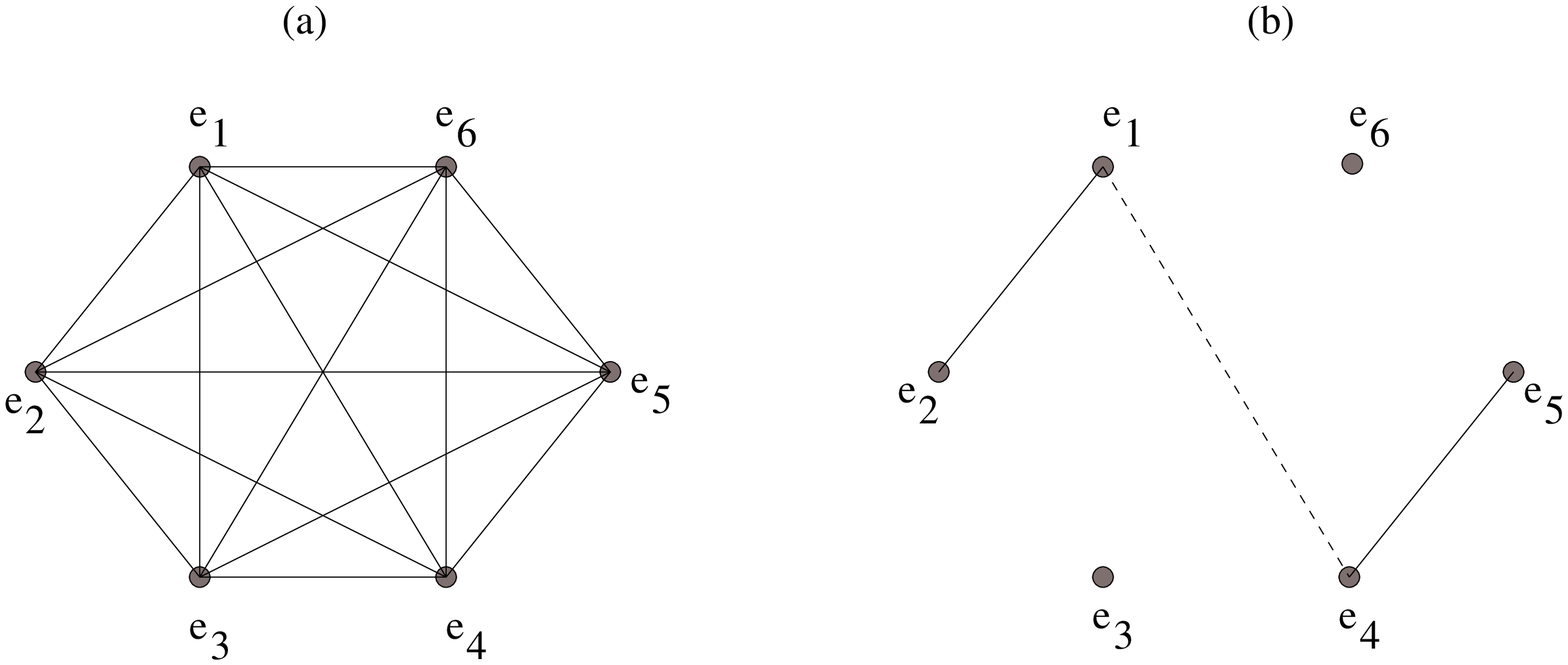}
}

It is helpful now to have an explicit look on the orientifold construction of
$SO(2k)$ and $Sp(2k)$ out of $SU(2k)$. Happily everything can be done in the
semiclassical
regime. The (classical) moduli space of an $SU(2k)$ gauge theory can be
represented by $2k$ points moving in a complex plane. Gauge symmetry
enhancement corresponds to colliding points (vanishing 0-cycles). Of course we
have a direct realization of this in terms of the endpoints of the fourbranes
moving in the v-plane! A typical situation for $k = 3$ is shown in figure
\cyclefig\ (a).
A simple basis of vanishing cycles is given by $e_1-e_2$, $e_2-e_3$, $e_3-e_4$,
$e_4 - e_5$ and $e_5-e_6$.
Their intersection form coincides with the Cartan matrix of $SU(6)$. The
orientifold projection selects now cycles that are odd under the space
reflection\foot{There is an additional minus for the vector excitation of the
string such that precisely the vectors are left invariant.}
$v \rightarrow -v$  \cyclefig\ (b).
For the case of an $SO(2k)$ projection a simple basis of vanishing cycles is
given by
$e_1-e_2-e_4+e_5$,  $e_2-e_3-e_5+e_6$ and $e_4-e_1-e_2+e_5$. Their intersection
form is proportional to the Cartan metric of $SO(6)$. For the case of an
$Sp(2k)$ projection a simple basis is given by
the first two of the former cycles and in addition $2(e_1-e_4)$. The
intersection form of these cycles is proportional to the metric on root space
of $Sp(6)$. The important point is that there is a subset of cycles left
invariant by both orientifold projections. However in the case of $SO(2k)$
these cycles give the long roots whereas the same cycles correspond to the
short roots in the case of $Sp(2k)$. If we fix the normalization of the common
roots to one, as is usual in physics, the length-squared of the long roots in
the symplectic group will be two. The one loop beta function coefficient is
given by a sum over the indices of the representations under which the various
fields in a gauge theory transform. These indices depend however upon the
normalization of the roots, e.g. for the adjoint $C_\theta = (\theta,\theta)
g^\vee$ where $\theta$ is the highest root and $g^\vee$ the dual Coxeter
number. This overall factor two can be absorbed into the definition of the
gauge
coupling of the $Sp(2k)$ theory.

Now we would like to derive the Riemann surface $F(t,v^2)=0$
that provides the exact solution for $\CN\!=\!2$ Yang-Mills theories with
symplectic gauge groups. From \easS, the curve should take the form
\eqn\ecS{t^2 +(v^2 B(v^2)+c) ~t + 1 =0.}
To reproduce the correct bending of the fivebranes at infinity,
$B(v^2)$ should be a polynomial even in $v$, of degree $2k$.
A general polynomial of this form will depend on $k+1$
parameters. The coefficient of $v^{2k}$ may be set to one by a
rescaling of the gauge coupling. However, this leaves us with one
parameter too many to describe the Coulomb branch of $Sp(2k)$ gauge
theory.
Let us analyze more carefully the behavior at
$v=0$. In the previous section we modeled the interaction
between the fivebranes and the orientifold as the deformation
induced on the fivebranes by the charge carried by the orientifold.
This deformation depends on the sign (and the magnitude) of the
charge. According to this, since now we are considering an orientifold
with the opposite charges, the deformation of the fivebranes should
point in the reverse direction.
\ifig\symplecfig{Behavior of fivebranes near an orientifold plane that
gives rise to $Sp(2k)$ gauge theory on fourbranes.}
{
\epsfxsize=3.2truein
\epsfysize=3.0truein\epsfbox{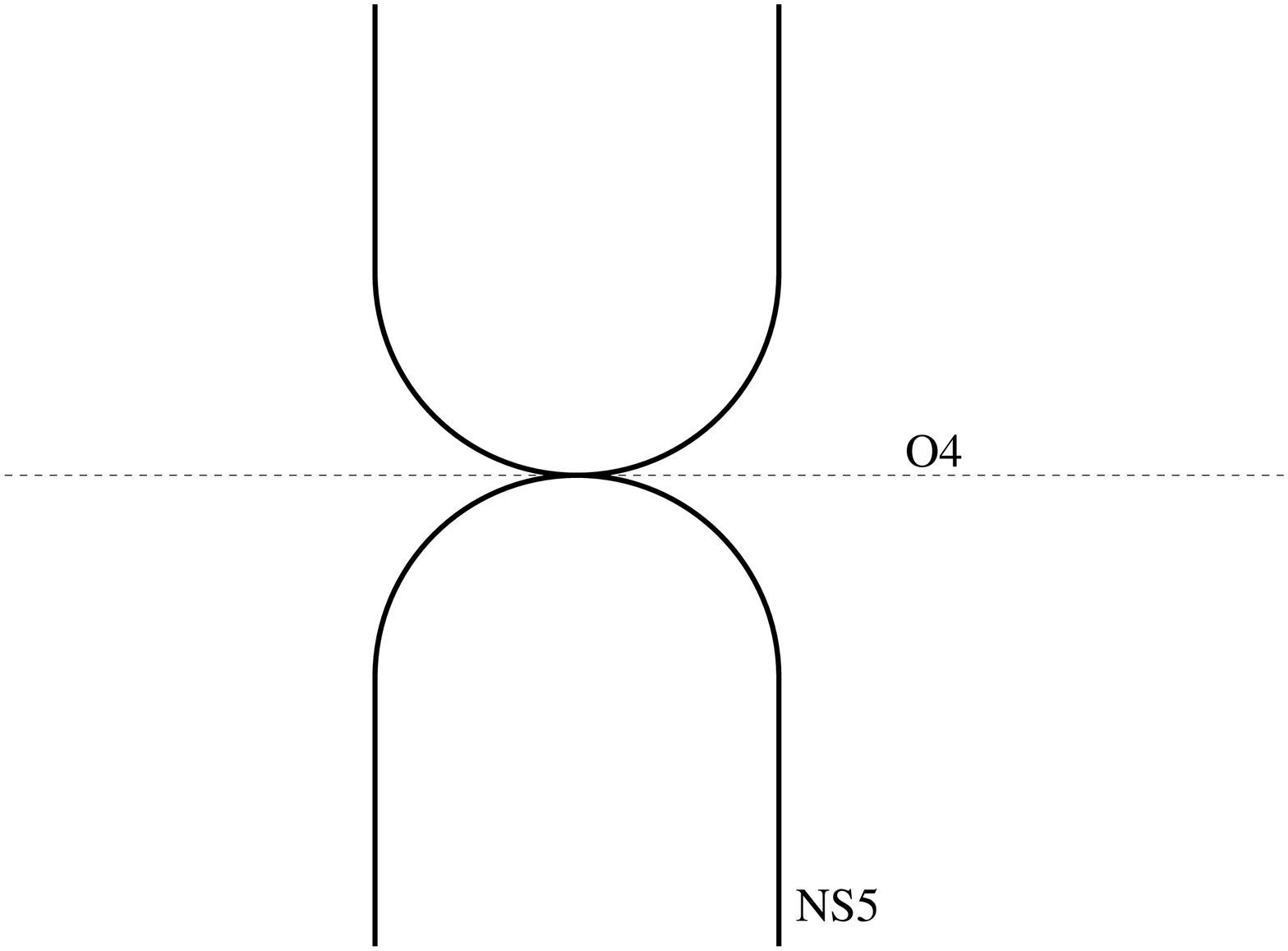}
}
Namely, instead of running to
$t=0,\infty$ at $v=0$ both fivebranes will be deformed towards
each other, as in \symplecfig. They will eventually meet at a central
point and the charges will cancel. When this happens
a single additional tube connects the two fivebranes. For this to
happen in a way which preserves the $v \to -v$, $t\to t$ symmetry we
must demand that \ecS\ has a double root at $v=0$.
This condition implies the curve is of the form\foot{There is a
misprint in the corresponding expression in \mw. It should read
$(z-\mu/z)^2 + x^{(2l+2)} + x^{(2l)} u_2 + \dots + x^2 u_{2l}$. Our
form of the curve agrees with this upon
setting $t=z^2$ and $\mu=1$. We thank N. Warner for a conversation on this 
point.} 
\eqn\eCS{t^2 + (v^2 B(v^2) - 2) ~t + 1 =0.}
Shifting $\tilde{t}=t+(v^2 B(v^2) - 2)/2$, this Riemann surface
can be rewritten as
\eqn\eS{\tilde{t}^2={(v^2 B(v^2) - 2)^2 \over 4} - 1,}
which is a double cover of the known curve solving $\CN\!=\!2$ pure gauge
theory
with
symplectic gauge group \as.

The degree of the polynomial is $4k+4$
which indicates $2k+2$ branchcuts. Due to the presence of the double point
we are however in a degenerate configuration where two branch-points coincide
at $v=0$ and two branchcuts melt into a single one. Therefore the genus of
the curve is $2k$ instead of $2k+1$. In the usual
representation of hyperelliptic curves the tubes
connecting the fivebranes are represented by branchcuts. Here we have one more
branchcut than expected from the number of fourbranes in the
semiclassical analysis. This means that
nonperturbative effects have generated the additional tube
between the fivebranes.
If we compute the discriminant of our curve for $Sp(4)$ with $u_2 = u$
and $u_4=w$ we find
\eqn\discriminant{ \Delta_{Sp(4)} = \Lambda^{12} w (u^2+4w)(-27\Lambda^{12} - 4
\Lambda^6 u^3  - 18  \Lambda^6 u w + u^2 w^2 + 4 w^3) .}
\ifig\branchfig{Branchcuts for the curve describing $Sp(4)$
Yang-Mills.}
{
\epsfxsize=3.0truein
\epsfysize=2.0truein
\epsfbox{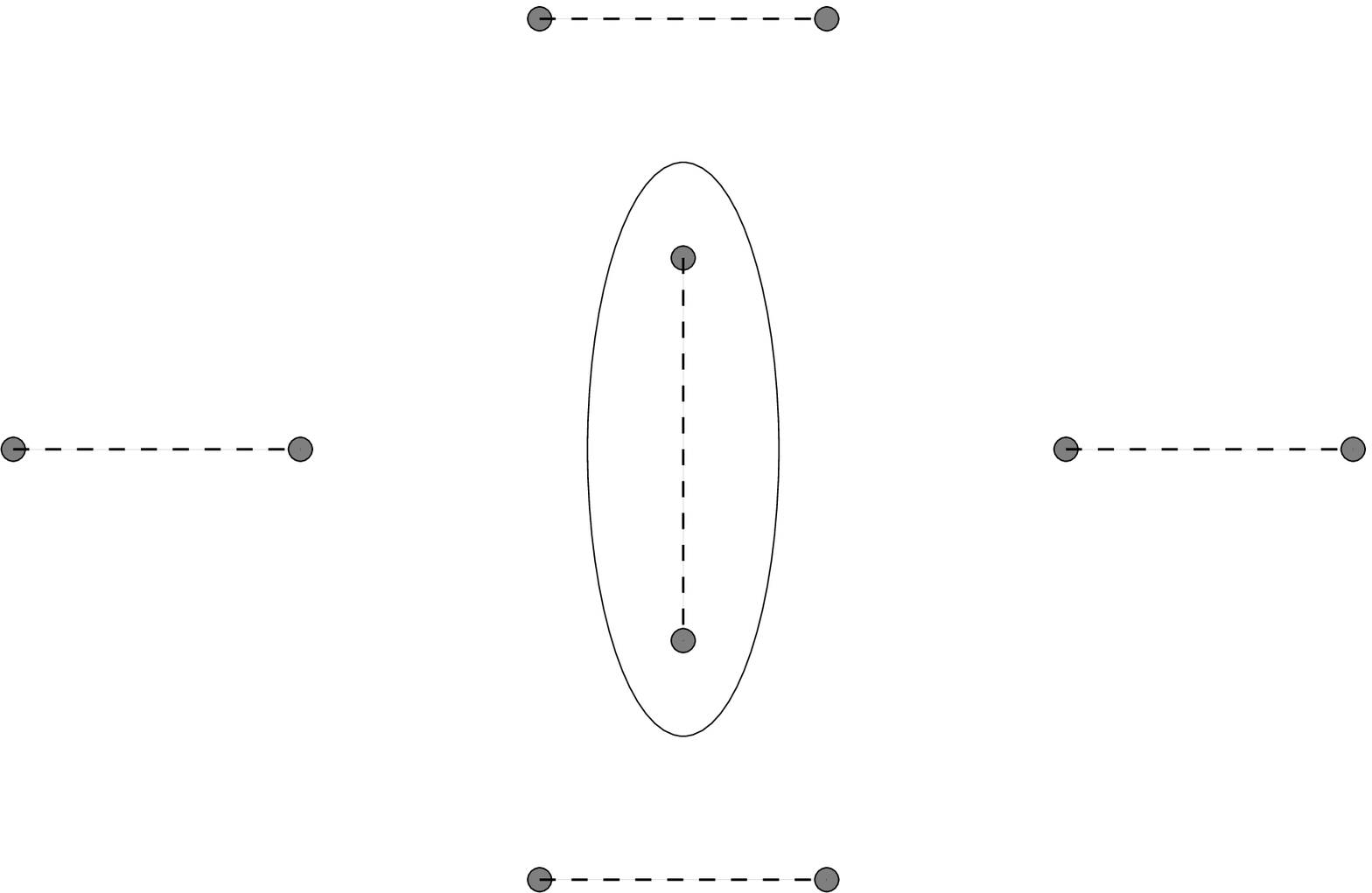}
}
Comparing this to what one gets from the curve in \as\ one notes that there is
an additional overall factor $w$ present in our case. This additional
singularity arises when the branchcut connecting the mirror images of
branchcuts in \branchfig,
shrinks to zero size. According to our argument in section two this should
not correspond to a physical singularity. Indeed the cycle around this
branchcut belongs to the even part of the homology under $v\rightarrow -v$. The
orientifold projection however selects the odd part. We find therefore another
type of apparent singularity in the curves for the symplectic gauge groups!

Let us stress the consistency of the picture we have obtained.
The fivebranes are pushed towards each other by the orientifold.
Once they meet, the configuration is stabilized since the
orientifold is seen by each fivebrane as a charge of
equal magnitude but opposite sign, i.e. $\pm2$.
Thus the region where the orientifold would have
charge $1$ shrinks to zero by nonperturbative effects.
We are left with an orientifold plane
which does not change nature even though it traverses the fivebranes.
The same situation was encountered in the previous subsection.
The orientifold plane in that case pushed the fivebranes off
each other at $v=0$. With nothing to stabilize them
they run out to $s=\pm\infty$, extending the region where the
orientifold has charge $-1$ all along the $s$-direction.

\subsec{$SO(2k+1)$ Gauge Groups}

In order obtain $SO(2k+1)$ gauge groups we will consider the same
brane configuration as in subsection (3.1) but with an additional
fourbrane lying over the orientifold. This new brane will be
taken to end on the fivebranes, in the same way as the $2k$ paired
fourbranes. The new single fourbrane is frozen at
$v=0$ since it does not have a mirror image.

In this case, the bending of the fivebranes at large $v$ is given
by
\eqn\easOO{t_i \sim v^{a_i} ~ , \quad a_1=-a_2=2k-1,}
because the orientifold and the additional fourbrane cancel charge
in the interval between fivebranes. Now the system of orientifold
plus fourbrane at $v=0$ is seen by the fivebranes as a charge
$+1$ on the left fivebrane and $-1$ on the right. Still these charges will
deform the fivebranes off each other to $t=0,\infty$,
as  happened for the $SO(2k)$ groups. The Riemann
surface that reproduces the expected behaviors at $v \rightarrow
\infty$ and $v=0$ is
\eqn\eCOO{v ~t^2 + B(v^2) ~t + v=0,}
with $B(v^2)$ as in \eB. Multiplying \eCOO\ by $v$ and redefining
$\tilde{t}= v t + B/2$, we get
\eqn\eOO{\tilde{t}^2 = {B(v^2) \over 4} - v^2}
which agrees with the spectral curve for $SO(2k+1)$ Yang-Mills theories
\ds.

While the Riemann surface in the form \eOO\ is invariant under
$v \rightarrow -v$, \eCOO\ is only invariant under the combined operation
$v \rightarrow -v$, $t \rightarrow -t$. In terms of the $(v,s)$
variables, this corresponds to modding out by the transformation
\eqn\eorOO{\eqalign{& v \rightarrow -v, \cr
&s \rightarrow s + i \pi\cr}}
combined with worldsheet parity reversal.

\subsec{Effects of Semi-Infinite Fourbranes}

We can now add mirror pairs of semi-infinite fourbranes to the left or
to the right of
the previous configurations. Whenever $v$  equals the position of such
a fourbrane $F(t,v)$ should have $t=\infty$ or $t=0$ as a
solution. Again this determines the curves uniquely. For the
orthogonal groups one finds
\eqn\somatter{\eqalign{&{F}_{SO(2k)}=\;v^2 t^2\prod_{i=1}^{N_f^L}( v^2 -m_i^2)
 + e B(v^2) t + f v^2 \prod_{j=1}^{N_f^R}( v^2 -m_j^2)=0\,, \cr
&{F}_{SO(2k+1)}=\; v t^2\prod_{i=1}^{N_f^L}( v^2 -m_i^2)
 +e B(v^2) t +f v \prod_{j=1}^{N_f^R}( v^2 -m_j^2)=0\, .}}
In the case of symplectic gauge groups we have one more
condition. The fivebranes should meet at $v=0$. We
therefore take as an ansatz
\eqn\spmatter{{F}_{Sp(2k)} = \prod_{i=1}^{N_f^L}( v^2 -m_i^2)
t^2 + e (v^2 B(v^2) - c) t + f \prod_{j=1}^{N_f^R}( v^2 -m_j^2)=0\,.}
Here we have allowed for arbitrary constants $e$ and $f$. As long as
the $\beta$-function of the theories is negative we can set these
constants to one by rescaling $t$ and $v$.
Once this is done, the constant $c$ is fixed by demanding that $F$ have a
double root at
$v=0$ and turns out to yield $c=2\prod_{i=1}^{N_f^L}\prod_{j=1}^{N_f^R}
m_i m_j $. In all cases it is easy to find a transformation that
brings these curves into the already known forms as summarized in
\kdp,
\eqna\trafos
$$\eqalignno{{\rm SO(2k):~~~~~~~~}& v^2 t\prod_{i=1}^{N_f^L} (v^2-m_i^2)
 = \tilde{t} - {B(v^2) \over 2}&\trafos a\cr
{\rm SO(2k+1):~~~~~~~~}& v t\prod_{i=1}^{N_f^L} (v^2-m_i^2)  = \tilde{t} -
{B(v^2) \over 2}&\trafos b\cr
{\rm Sp(2k):~~~~~~~~}& t \prod_{i=1}^{N_f^L} (v^2-m_i^2)  = \tilde{t} - {v^2
B(v^2) - c \over 2}&\trafos c\cr}
$$

Let us now ask what happens if we place a single semi-infinite
fourbrane on top of the orientifold plane to the right of a
configuration with
symplectic gauge group. Such a configuration
gives only half as many states as are needed to form a hypermultiplet in
the fundamental of the $Sp$ gauge group. These states will form a half
hypermultiplet. A mass term is not possible for half hypermultiplets. In our
brane
configuration such a term would
correspond to moving the semi-infinite fourbrane off the
orientifold. Naively applying the same arguments as in the previous
cases we find  that the
curve describing this configuration is
\eqn\halfhyper{F_{1/2 hyper} = t^2 + v^2 B(v^2) t + v \,.}
However, this is not consistent since the curve is not symmetric under
a $\IZ_2$ symmetry taking $v\to -v$.
The only way to achieve this now is to take $|v|$ instead of $v$
in the last term of \halfhyper. Clearly this destroys the complex
structure and therefore breaks supersymmetry.
It is indeed well known
that $\CN\!=\!2$ theories with symplectic gauge groups and an odd
number of half
hypermultiplets suffer from Witten's global anomaly \edanom.
We interpret the inconsistencies of the
curve for the brane configurations corresponding to these gauge
theories as a manifestation of this global anomaly. That we can
actually detect this through D-brane considerations shows how powerful
this approach is.

In the case of vanishing $\beta$-function we have to take into account
that we can absorb only one of the two constants $e$ and $f$ by
scalings. The remaining parameter can be adjusted arbitrarily and
determines the UV-value of the gauge coupling. Translated into our
brane configuration this states that the positions of the fivebranes
are well-defined at large $v$. More precisely, one finds $t \sim
\lambda_\pm v^a$ where $a=2k$ or $a=2k+2$ for orthogonal and
symplectic gauge groups respectively and $\lambda_\pm$ are the roots
of $y^2 + e y + f$. This is the same structure as in \supered\ and
following the arguments of that paper
the duality group is $\Gamma_0(2)$ in our cases as well.

The cases of positive $\beta$-function can be described as follows. For
the gauge groups of the form $SO(2k)$ after using the transformation
\trafos{a} the curves are of the form
\eqn\somestuff{
\tilde{t}^2 = e^2 B^2(v^2) - f v^4 \prod_{i=1}^{N_f} (v^2- m_i^2)\,.
}
For $N_f > 2k -2$ the asymptotic behavior of the fivebranes shows that
they become parallel at large $v$.
It is not changed if we add terms in $B(v^2)$\foot{Here we do not admit
terms that would break the symmetry $v\rightarrow -v$.}. The highest
term we can add in this manner is of order $2k' = N_f+2$ if $N_f$ is
even and of order $2k' = N_f +1$ if $N_f$ is odd. If $N_f$ is even the
resulting curve
is the one of the $SO(N_f+2)$ theory with vanishing $\beta$-function.
The physical interpretation is clear: the UV strongly coupled theory is
embedded in the theory with smallest gauge group of the same family
with finite UV-behavior. If $N_f$ is odd there is no way the
theory can be deformed into a usual gauge theory without changing the
asymptotic behavior. The gauge group within the same family can be at
most enlarged to $SO(N_f+1)$. In fact the positions of the fivebranes
go as $t_\pm = \pm \sqrt{-f} v^{N_f+2}$. The gauge coupling is
proportional to $\log(t_+/t_-)$. The theory seems to
flow to a genuinely strongly coupled fixed point with no adjustable free
parameter. This is essentially the same behavior as found in \supered\ for
unitary gauge groups.

For symplectic gauge groups the story is similar. Here we need $N_f >
2k+2$. Again, if $N_f$ is even this theory flows to the one with
vanishing $\beta$-function and $N_f$ flavors. For $N_f$ odd there is a
similar strongly coupled fixed point.
For gauge groups of the form $SO(2k+1)$ the behavior is just the
opposite. If $N_f$ is odd one can embed the theory into a UV-finite one
with gauge group $SO(N_f+2)$. The strongly coupled fixed point appears
for $N_f$ even.

\subsec{Product Groups}

In this subsection we want to derive the exact solution for
gauge theories whose gauge group is a product of orthogonal
and symplectic groups. We start with the case where the gauge group
involves $SO(2k)$ and $Sp(2k)$ groups only.

\bigskip
\noindent
{\it Gauge Groups of the form $\cdots \times SO(2k_{\alpha -1}) \times
Sp(2k_{\alpha}) \times SO(2k_{\alpha +1}) \times \cdots$ }

In order to get product groups we need to
consider configurations with more than two fivebranes.
We will consider a chain of $n+1$ fivebranes, with
$2k_{\alpha}$ and $2k_{\alpha+1}$ fourbranes ending respectively
on the left and right of the $\alpha$-th fivebrane
($\alpha= 0,\cdots,n$). We will assume that there are no semi-infinite
fourbranes on the ends of the configuration, therefore
$k_0=k_{n+1}=0$. As before an orientifold plane will traverse
the whole configuration at $v=0$.
In the Type \IIA\ string picture each time the
orientifold crosses a fivebrane it changes nature.
The product group structure we obtain is
\eqn\eprod{\cdots \times SO(2k_{\alpha -1}) \times
Sp(2k_{\alpha}) \times SO(2k_{\alpha +1}) \times \cdots .}
We also get $n-1$ half hypermultiplets transforming as
$({\bf 2k_{\alpha}}, {\bf 2k_{\alpha+1}})$, where
${\bf 2k_{\alpha}}$ denotes the fundamental representation
of the corresponding orthogonal or symplectic gauge group,
$G_{\alpha}$. With respect to each $G_{\alpha}$ there is always an even
number of half hypermultiplets present.

In the following we want to consider theories with negative or zero
$\beta$ functions.
As in the previous sections, we can read the one-loop $\beta$ function
coefficient from the bending of the fivebranes at large $v$
\eqn\easP{t_{\alpha} \sim v^{a_{\alpha}} \;\;,
{}~~~~~~~~~~a_{\alpha}=2k_{\alpha+1} - 2k_{\alpha} - 2 \omega_{\alpha},}
where $\omega_{\alpha}$ is the charge of the orientifold on the
left of the $\alpha$-th fivebrane. From \gaugecoupling,
the one-loop $\beta$-function coefficient $b_{0,\alpha}$ for the gauge
group $G_{\alpha}$
is proportional to $a_{\alpha} - a_{\alpha -1}$. Thus the condition
$b_{0,\alpha} \leq 0$ implies
\eqn\eorP{a_0 \geq a_1 \geq \cdots \geq a_n.}

The exact solution of this model will be given in terms
of a Riemann surface $F(t,v^2)=0$, where now $F$ is a polynomial
of order $n+1$ in $t$
\eqn\ecP{P_0 (v^2) ~t^{n+1} + P_1 (v^2) ~t^{n} + \cdots + P_{n+1} (v^2)
=0.}
The relation \eorP\ allows one to determine the degree of the
polynomials $P_i (v^2)$. This can be seen by rewriting \ecP\ as
\eqn\ecauxP{P_0(v^2) \prod_{\alpha = 0}^{n} (t - t_{\alpha}(v^2))
=0,}
where $t_{\alpha}$ are now rational functions of $v^2$ with the
asymptotic behavior \easP. Therefore the degree of $P_i$ is
\eqn\edegP{p_i = \sum_{j=0}^{i-1} a_{\alpha} + p_0=
2k_{i} - (1-(-1)^i) ~\omega_0 + p_0,}
with $p_0$ the degree of $P_0$. Since we assumed that there
are no semi-infinite fourbranes on either end of the configuration,
$P_0$ depends only on how the orientifold deforms the leftmost fivebrane.
If $\omega_0=1$ the first factor in the chain is an orthogonal group,
in which case the orientifold will push the first fivebrane
to $t=\infty$. Using the results of subsection (3.2) we have then
$p_0=2$.
If $\omega_0=-1$ the first group of the chain is symplectic.
The orientifold will deform the first fivebrane towards the next one.
In this case $t$ remains finite at $v=0$ and therefore $p_0=0$.
We can now write \edegP\ in the simple form
\eqn\eDegP{\eqalign{& p_i = 2 k_i +1+(-1)^i ,
{}~~~~~~~~~~~~\omega_0=1, \cr
& p_i = 2 k_i +1+(-1)^{i+1},
{}~~~~~~~~~\omega_0=-1. \cr}}

It is convenient to explicitly mention the four different cases we
can obtain. If there is an even number
of fivebranes the first and last groups of the chain will belong to
the same series. More precisely, for $\omega_0=1$ we get the
chain $SO(2k_1) \times Sp(2k_2) \times \cdots \times SO(2k_n)$.
For $\omega_0=-1$ we will obtain $Sp(2k_1) \times SO(2k_2) \times
\cdots \times Sp(2k_n)$. If there is an odd number of fivebranes
the first and last groups will differ. When $\omega_0=1$ we
derive the chain $SO(2k_1) \times Sp(2k_2) \times \cdots \times
Sp(2k_n)$. On the contrary, when $\omega_0=-1$ we will find
$Sp(2k_1) \times SO(2k_2) \times \cdots \times SO(2k_n)$.
In order to show the structure of the solution we will consider
this last case, i.e. $\omega_0=-1$ and $n+1$ odd.
Knowing this case, the others can be obtained in
a straightforward way.

Using \eDegP, the Riemann surface \ecP\ for $\omega_0=-1$
and $n+1$ odd looks like
\eqn\ecurP{t^{n+1} + (v^2 B_1 (v^2) + c_1) ~t^n + B_2 (v^2) ~t^{n-1}
+ \cdots +  v^2 + c_{n+1}=0.}
The functions $B_i$ are generic polynomials even in $v$ of degree
$2 k_{\alpha_i}$. The coefficients of the highest order term of $B_i$
determine the asymptotic behavior of the fivebranes and should be
interpreted as coupling constants \supered. One of them can be
eliminated by rescaling $v$. The remaining coefficients in the
$B_i$ can be interpreted as the Casimirs of $G_{\alpha_i}$.
$c_1,c_3,\cdots,c_{n+1}$ are constants
that can be determined from the expected behavior at $v=0$.
In the case we are now considering, the orientifold pushes
the rightmost fivebrane off to infinity. Therefore $t=0$ must be
a solution of \ecurP\ at $v=0$, which fixes $c_{n+1}=0$.
The analysis of subsection (3.3) tells us that, with the
exception of $t=0$, all the other roots of \ecurP\ at $v=0$
should be double roots
\eqn\ecurzeroP{t^n + c_1 ~t^{n-1} + u_{2k_2}^{(2)} ~t^{n-2}
+ \cdots + u_{2k_n}^{(n)} = \prod_{i=1}^{n/2} (t-t_i)^2,}
where $u_{2k_\a}$ is the square of the exceptional Casimir of
order $k_\a$
of $G_{\alpha}=SO(2k_{\alpha})$. Condition \ecurzeroP\
completely determines the constants $c_i$ in terms of
$u_{2k_\a}$.

Contrary to the case of the $SU(k)$  product groups considered in \supered,
in our case we can not introduce bare mass parameters for the
$({\bf 2k_{\alpha}}, {\bf 2k_{\alpha+1}})$ half-hypermultiplets.
Bare mass parameters correspond to a non-zero constant in \finite\
which is not allowed by the $\IZ_2$ symmetry $v \rightarrow -v$
of our configurations. However this fact does not represent a lack
of generality of the brane construction, since it can equally be
derived from pure gauge theory considerations.

It is enough to analyze the case $G=Sp(2k_1) \times SO(2k_2)$
and one half-hypermultiplet $({\bf 2k_1}, {\bf 2k_2})$.
Let us use $\CN=1$ superspace notation and represent
the half-hypermultiplet by $X_{a}^{i}$, with $a=1,\cdots,2k_1$
and $i=1,\cdots,2k_2$. In this example the flavor group has
been completely gauged. The only way to write a gauge-invariant
mass term is
\eqn\emass{m X_{a}^{i}  X_{b}^{j} \delta_{i j} J^{a b},}
where $J^{a b}$ and $\delta_{i j}$ are the invariant matrices
associated to symplectic and orthogonal groups
respectively. Since $J^{a b}=-J^{b a}$ is antisymmetric
while $\delta_{i j}$ is symmetric, the bare mass term \emass\ is
identically zero. However the $4 k_1 k_2$ hypermultiplets
$X_{a}^{i}$ can acquire masses by turning on Higgs expectation
values. This corresponds to the $\CN=1$ superpotential
\eqn\superP{W= X_{a}^{i}  X_{b}^{j}  \phi_{Sp}^{a b}
\delta_{i j} + X_{a}^{i}  X_{b}^{j}  \phi^{SO}_{i j}
J^{a b},}
where $\phi_{Sp}^{a b}$ and $\phi^{SO}_{i j}$ are
chiral $\CN=1$ fields in the adjoint representation of
$Sp(2k_1)$ and $SO(2k_2)$ respectively.
In all the cases we treat in this subsection a maximal
subgroup of the flavor group has been gauged, therefore
the same field theory argument implies that bare mass
parameters are not allowed.

Let us analyze in more detail the curve for $G=Sp(2k_1) \times SO(2k_2)$.
The Riemann surface \ecurP\ for this case is
\eqn\eCP{t^3 + (e v^2 B_{Sp}(v^2) - 2 \prod_{i=1}^{k_2} a_{SO,i}) ~t^2
+ B_{SO} (v^2) ~t + v^2=0.}
We choose to eliminate the coefficient of $v^{2k_2} t$.
Both polynomials $B_{Sp}$ and $B_{SO}$ can be written as
\eqn\eBP{B_G (v^2) = \prod_{i=1}^{k_i} (v^2 - a_{G,i}^2),}
where $\pm a_{G,i}$ represent the (classical) positions of the
fourbranes in the $v$-plane. The curve \eCP\ is gauge invariant only
because for
orthogonal groups of the form $SO(2k)$ the product
$\prod_{i=1}^{k} a_i$ is a gauge invariant quantity!

We want now to take the leftmost or the rightmost fivebrane to
$\pm \infty$ and recover from \eCP\ the curves solving even
orthogonal and symplectic gauge groups with matter in
the fundamental representation. This will provide a
check of the curves we have proposed.
If we send the third fivebrane to infinity, the $SO(2k_2)$
gets effectively frozen since its classical gauge coupling
is sent to zero by this process. The $2k_2$
fourbranes with boundary on the second and third fivebranes
become now semi-infinite fourbranes. In this way
we reduce to $Sp(2k_1)$ gauge group, plus $k_2$ $\CN\!=\!2$
hypermultiplets in the fundamental representation.
In \eCP\ we have set the scales to one, alternatively
we could have considered
\eqn\eCPS{t^3 + (ev^2 B_{Sp}(v^2) - 2 \prod_{i=1}^{k_2} a_{SO,i}) ~t^2
+ B_{SO} (v^2) ~t + \Lambda v^2=0.}
By sending $\Lambda \rightarrow 0$ we effectively take
the third fivebrane to infinity, i.e. $t=0$. In this
limit $\prod_{i=1}^{k_2} a_{SO,i}$ behaves as the product of the
bare masses of the $k_2$ fundamental hypermultiplets.
Using \eBP, the curve \eCPS\ becomes
\eqn\eCSM{t^2 + (e v^2 B_{Sp}(v^2) - 2
\prod_{i=1}^{k_2} a_{SO,i} ) ~t + \prod_{i=1}^{k_2} (v^2 - a_{SO,i}^2)=0.}
The factor $e$ can now be set to one by scaling $v$ and $t$
appropriately and the expression coincides with the curves for
symplectic gauge groups.

In the same way, if we send the first fivebrane to $- \infty$,
we will reduce the gauge group to $SO(2k_2)$, plus $k_1$
hypermultiplets in the fundamental representation.
As before we can introduce a scale $\Lambda$ in \eCP\ which
corresponds to shifting the position of the first fivebrane.
However we must be careful to do it in a way that preserves
the condition \ecurzeroP. This can be achieved by
\eqn\eCPO{\Lambda^2 t^3 + (e v^2 B_{Sp}(v^2) - 2 \Lambda u_{2k_2}) ~t^2
+ B_{SO} (v^2) ~t + v^2=0.}
The limit $\Lambda \rightarrow 0$ corresponds to sending
the first fivebrane to $s=- \infty$, i.e. $t=\infty$.
Using again \eBP\ and scaling $v$ and $t$, \eCPO\ allows us  to
recover the solution for the orthogonal groups
\eqn\eCPOi{\prod_{i=1}^{k_1} (v^2 - a_{i}^2)  ~t^2
+ B_{SO} (v^2) ~t + v^2=0.}

The solutions for gauge groups of the form $SO(2k_1)\times Sp(2k_2)
\times \cdots SO(2k_n)$ can be worked out in a similar way and take the form
\eqn\sospsoeven{ F(t,v^2) = v^2 t^{n+1} + B_1(v^2) t^n + (v^2 B_2(v^2)
+ c_1) t^{n-1} + \cdots + B_n(v^2) t + v^2.}
Here $n+1$ is an even integer. At $v=0$ this has $t=0$ and $t=\infty$ as
solutions. Again we demand the other zeroes at $v=0$ to be double
points. This fixes the $(n-1)/2$ constants $c_i$.

For gauge groups of the form $Sp(2k_1)\times SO(2k_2) \times \cdots
Sp(2k_n)$
one finds
\eqn\spsosp{F(t,v^2) = t^{n+1} + (v^2 B_1(v^2) + c_1) t^n + B(v^2)
t^{n-1} + \cdots + (v^2 B(v^2) + c_n) t + 1 .}
Again $n+1$ is an even integer and the $( n+1)/2 $ constants $c_i$
are fixed by demanding that we double points only at $v=0$.

\bigskip
\noindent
{\it Gauge Groups of the form $\cdots \times SO(2k_{\a-1}+1)\times
Sp(2k_{\a}) \times SO(2k_{\a+1}+1) \times \cdots$}

It is also possible to have gauge groups consisting of factors of the
form $SO(2k_1+1)\times Sp(2k_2) \times SO(2k_3+1)$. In this case one has
two massless half hypermultiplets in the $Sp$ groups and this gives a
consistent theory. However, we still cannot give these half
hypermultiplets a mass. We can understand this by taking the view of
gauging the global flavor symmetry of a $\CN\!=\!2$ gauge theory with
symplectic gauge group. The flavor symmetry is always $SO(2N_f)$ with
$N_f$ being the number of hypermultiplets. The brane configuration
corresponds to the case when one gauges only a $SO(2k_1+1)\times
SO(2k_3+1)$ subgroup of $SO(2N_f)$, where of course $(k_1+k_3+1) =
N_f$. Giving masses to the hypermultiplets corresponds now to turning on vev's
of the Higgs fields $\Phi_{SO}$ in the adjoint of $SO(2k_1+1)$ and $SO(2k_3+1)$
respectively. Since $\Phi_{SO}$ has to lie in the Cartan subalgebra it
is of the form $\Phi = {\rm diag}(a_1\, i\sigma_2\,, \cdots\,,a_{k_i}\,
i\sigma_2,
0)$ where we used the usual form of the Pauli matrix $\sigma_2$. In
$\CN=1$ superspace the
superpotential of such a theory can be written as
\eqn\supo{ W = X^i_a X^j_b \Phi^{ab}_{Sp} \delta_{ij} +  X^i_a X^j_b
{J}^{ab} \Phi_{ij}^{SO_1} +  Y^i_a Y^j_b \Phi^{ab}_{Sp}
\delta_{ij} + Y^i_a Y^j_b {J}^{ab} \Phi_{ij}^{SO_2}\,.}
The fields transforming in the fundamental of the first and second $SO$ factor
are denoted with $X$ and $Y$ respectively, $J^{ab}$ is the
symplectic metric. In
this way one sees directly that the two half hypermultiplets remain
massless.

It is easy to generalize the discussion of the previous case to this
situation. Without going into the details we state the resulting form
of the curves for the gauge group being $SO(2k_1+1)\times Sp(2k_2)
\times SO(2k_3+1) \times \cdots \times SO(2k_n +1)$
\eqn\sospso{ F(t,v) = v t^{n+1} + B_1(v^2) t^{n} + v B_2(v^2)
t^{n-1} + B_3(v^2) t^{n-2} + \cdots + v = 0\, ,}
where $n+1$ is an even integer.
The curve respects the symmetry $v\rightarrow -v$, $t \rightarrow
-t$. We find that the number of relevant parameters
in \sospso\ matches precisely the number of Casimirs and couplings of
the gauge group factors.

When deriving the curves associated with symplectic groups and product
groups containing symplectic and even orthogonal groups we had to use a crucial
ingredient. We asked that, except for possible solutions $t=0$ or
$t=\infty$, the roots of $F(t,0)=0$ should be double roots. This arose
from requiring that when the fivebranes are deformed towards each
other, they meet to form a single tube.
Although for products of odd orthogonal and
symplectic groups we do not require additional
restrictions to fix the unique form of the curves \sospso, it should
be pointed out the same picture holds in these cases as well.
At $v=0$ \sospso\ has as solutions
$t=0,\infty$ and the roots of
\eqn\eroo{ u_{2k_1}^{(1)} ~t^{n-1} + u_{2k_3}^{(3)} ~t^{n-3} +
\cdots + u_{2k_n}^{(n)}= u_{2k_1}^{(1)} \prod_{i=1}^{(n-1)/2}
(t^2 - t_{i}^2)=0.}
The solutions of \eroo\ are of the form $t=\pm t_i$. However due to
the $\IZ_2$ symmetry $v\rightarrow -v$, $t \rightarrow
-t$ these correspond to double points in the quotient space.

\newsec{Including Six-Branes}

Now we consider the addition of sixbranes to the configurations
of fourbranes and fivebranes previously discussed. The
sixbranes are extended in the $x^0,\cdots, x^3$ and $x^7, x^8$ and $x^9$
directions. Each sixbrane is accompanied by its image under
the
action of the orientifold symmetry.

\ifig\fsixbrane{A configuration of fourbranes, fivebranes and sixbranes.}
{
\epsfysize=2.8in\epsfxsize3.2in\epsfbox{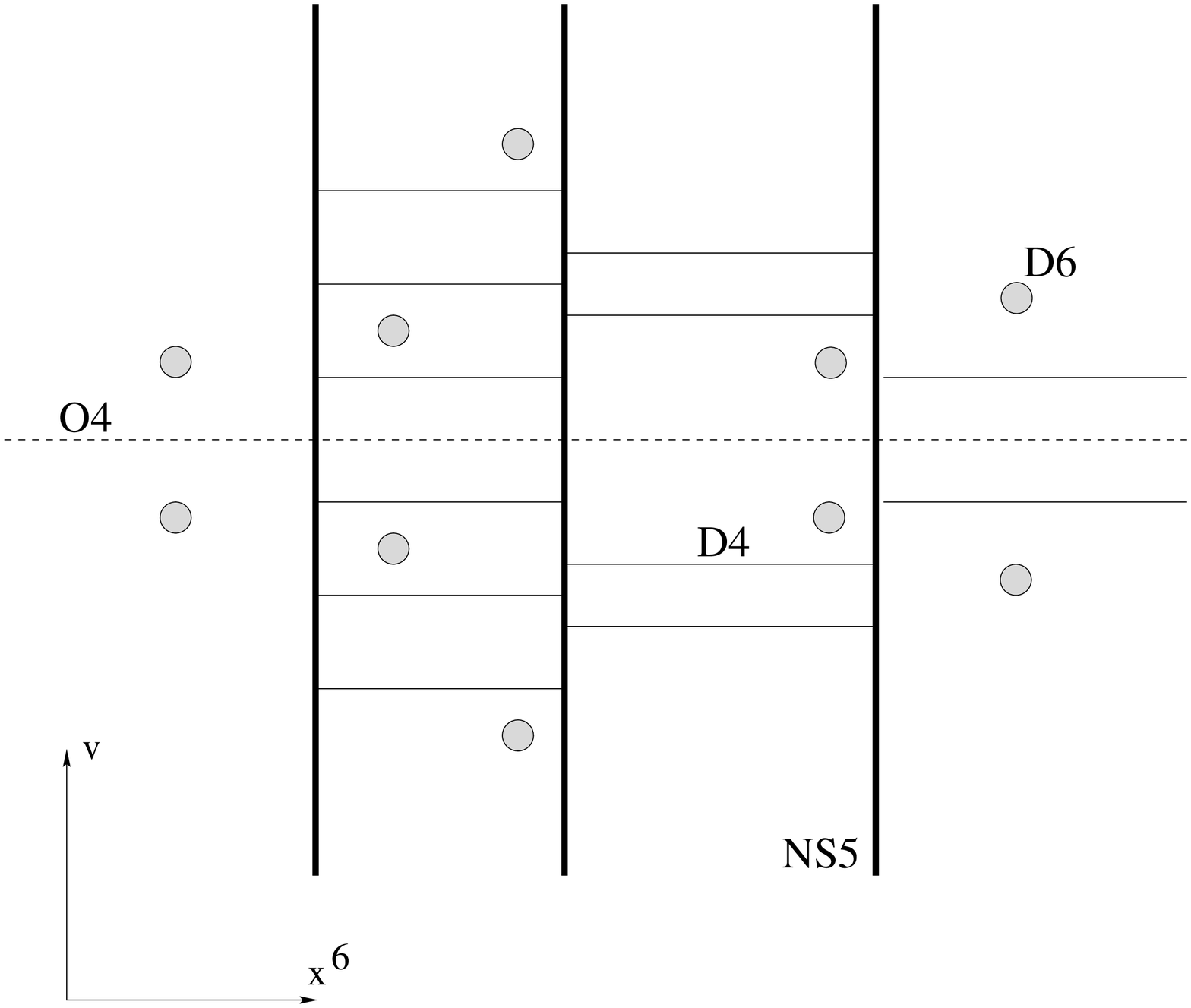}}

An example of such a configuration is shown in \fsixbrane.
Let $d_\alpha$ denote the number of pairs of sixbranes between the
$(\alpha\!-\!1)$-th and $\alpha$-th fivebrane. Open strings
running between the sixbranes and fourbranes will give rise
to $d_\alpha$  additional hypermultiplets in the fundamental
representation of the gauge group.

A configuration of parallel sixbranes in M-theory corresponds to the
product of a multi-Taub-NUT metric with flat $\IR^7$ space \ptowns. The
multi-Taub-NUT metric \hawking\ takes the form
\eqn\multinut{
ds^2 = {V \over 4} d {\vec r}^2 + {V^{-1} \over 4} (d \tau + \vec
\omega \cdot \vec r)^2~,
}
where
\eqn\vfunc{
\eqalign{
V&= 1+ \sum_{a=1}^d {1\over |\vec r -\vec x_a |}~, \cr
\vec \nabla \times \vec \omega &= \vec \nabla V ~.\cr}
}
The $\vec x_a$ are the positions of the sixbranes. It should be
noted this eleven-dimensional solution is nonsingular. This will
allow us to use the properties of low-energy M-theory to solve the
low-energy physics of the gauge theory.

To construct the gauge theory curves we do not need the full details
of the metric \multinut\ but only a description of the space in one
of its complex structures \supered. Such a description has been constructed in
\gibbons\
\eqn\nutspace{
yz= P(v)=\prod_{a=1}^d (v^2-e_{a}^2)~.
}
This four-dimensional space replaces the flat $\IR^3 \times S^1$
of the $x^4, x^5, x^6, x^{10}$ directions of the previous sections.

The fourbranes behave as before under the action of the orientifold
symmetry $\Omega$. Depending on the choice $\Omega^2=\pm$ we obtain orthogonal
($\Omega^2=1$) or symplectic gauge groups ($\Omega^2=-1$). $\Omega^2$
acts with the opposite sign on sixbranes, so the flavor symmetry group
that arises is symplectic for ($\Omega^2=1$) and orthogonal for
($\Omega^2=-1$). This restriction on the flavor symmetry groups, as we
have seen before, is
familiar from the field theory viewpoint.

The Seiberg-Witten differential for these gauge theories may be
constructed following section 2. Now the Riemann surface is embedded
in a curved spacetime. The area of the spatial component of the
membrane
worldvolume $D$ will now be given by the formula
\eqn\susysixa{
V \sim \biggl| \int_D {dv ~dz \over {\partial W/\partial y}}\biggr| =
\biggl|\int_{\partial D}  {v ~dz \over z} \biggr|~,
}
where we have introduced $W(v,y,z)= zy-P(v)$. The Seiberg-Witten
differential is then $\lambda = v ~dz/z$.

\subsec{ $SO(2k)$ with $d$ Fundamentals}

Let us first consider the case when we have a pair of fivebranes
and $d$ pairs of sixbranes. We wish to impose the condition that there
are no semi-infinite fourbranes to the left or to the right of the
fivebranes. The orientifold plane induces fourbrane
charge on the fivebrane, as previously discussed.
This means that as $v\to 0$, $y$ must have a solution that goes as
$1/v^2$ and one that goes as $v^2$. The curve must also be invariant
under the $\IZ_2$ symmetry $v\to -v$, with $y$ and $z$ invariant.
These conditions restrict the curve
to the form
\eqn\sosixcur{
v^2 y^2 + B(v^2) y + v^2 C(v^2) = 0~.
}

Substituting $z= P(v^2)/y$, where $P(v^2) = \prod_{a=1}^d (v^2-e_a^2)$
(remembering we have pairs of sixbranes positioned at $v=e_a$ and
$v=-e_a$)
we obtain
\eqn\sosixz{
C(v^2) v^2 z^2 + B(v^2) P(v^2) z + v^2 P(v^2)^2 = 0~.
}
In order that $z$ have only a solution that goes as $1/v^2$ and $v^2$
as $v\to 0$ we require that $BP$ and $P^2$ are divisible by $C$.
Taking the $e_a$'s to be distinct, the solution for $C$ is
\eqn\sosixc{
C=f \prod_{a=1}^{i_0} (v^2-e_a^2)^2 \prod_{b=i_0+1}^{i_1} (v^2-e_a^2)~,
}
where $i_0$ and $i_1$ are integers and $f$ is a constant. Following \supered,
$i_0$ will be the number of pairs of sixbranes to the left of the fivebranes,
and $i_1-i_0$ will be the number of pairs of  sixbranes between the fivebranes.
The solution for $B$ takes the form
\eqn\sosixb{
B= \tilde B(v^2) \prod_{a\leq i_0} (v^2-e_a^2)~,
}
for some polynomial $\tilde B$.

Defining $\tilde y= y/\prod _{a\leq i_0} (v^2-e_a^2)$ the curve
\sosixcur\ becomes
\eqn\sosixcc{
v^2 \tilde y^2+ \tilde B(v^2) \tilde y + f v^2 \prod_{a=i_0+1}^{i_1}
(v^2-e_a^2)=0~.
}
When $\tilde B$ is a polynomial of order $k$ in $v^2$, this is the
curve
describing $SO(2k)$ gauge group with $d$ flavors of fundamental
matter.

\subsec{$Sp(2k)$ with $d$ Fundamentals}

The above arguments carry over in a straightforward way to the
$Sp(2k)$ case. Now the orientifold projection gives rise to an
opposite
charge for the orientifold plane. $y$ should therefore have no
solutions that go to zero or infinity for finite $v$. Imposing the
$\IZ_2$ symmetry  $v\to -v$, with $y$ and $z$ fixed, yields the curve
\eqn\spsixcur{
y^2 + B(v^2) y + C(v^2) =0~.
}
Substituting in $z=P/y$, with $P$ as defined above, we find
\eqn\spsixz{
C(v^2) z^2 + B(v^2) P(v^2) z + P(v^2)^2 =0~.
}
In order that $z$ has no solutions that go to zero or infinity for
finite $v$ corresponding to semi-infinite fourbranes we must have that
$C$ divides $BP$ and $P^2$. The solution of these conditions is
\sosixc\ and \sosixb. Defining $\tilde y$ and $\tilde B$  as before we
obtain the
equation
\eqn\spsixcc{
\tilde y^2 + \tilde B \tilde y + f \prod_{a=i_0+1}^{i_1}
(v^2-e_a^2)=0~.
}

As in the previous examples of $Sp(2k)$ gauge theories considered
above,
we must impose the condition that $y$ has a double root at $v = 0$.
This fixes
\eqn\spsixbo{
\tilde B(0) = 2 f^{1/2} \prod_{a=i_0+1}^{i_1} i e_a~.
}
$\tilde B(v^2)$ may then be written $v^2 B'(v^2) + \tilde B(0)$.
Assuming $B'$ is a polynomial in $v^2$ of order $k$, we find upon
substituting back into \spsixcc\ the familiar
curve for $Sp(2k)$ with $i_1-i_0$ fundamentals.

\subsec{$SO(2k+1)$ with $d$ Fundamentals}

The only essential difference between this case and the $SO(2k)$ case
is that now we have an additional fourbrane lying on the
orientifold plane frozen at $v=0$ and stretched between the
fivebranes.
This changes the asymptotic behavior of the fivebranes, so that
now we must have a solution $y \sim 1/v$ and $y \sim v$ as $v \to 0$.
The $\IZ_2$ orientifold symmetry now acts as $v\to -v$, $y\to -y$, and
$z\to -z$. These constraints imply the curve takes the form
\eqn\soosixcur{
v y^2 + B(v^2) y + v C(v^2) =0~.
}
Following the same procedure as above, we can solve for $B$ and $C$
and we obtain the same answer \sosixc\ and \sosixb.
With $\tilde y$ defined as before, the equation for the curve is
\eqn\soosixcc{
v \tilde y^2 + \tilde B \tilde y + v \prod_{a=i_0+1}^{i_1}
(v^2-e_a^2)=0~.
}
Taking $\tilde B(v^2)$ to be a polynomial of degree $k$ in $v^2$, we
obtain the curve for $SO(2k+1)$ with $i_1-i_0$ fundamentals.

\subsec{Product Gauge Groups}

We now generalize to the models with $n+1$ fivebranes. Depending on
the orientifold projection, we can obtain alternating products
of $Sp(2k)$ with $SO(2k')$ gauge groups, or $Sp(2k)$ with $SO(2k'+1)$
gauge groups.

\bigskip\noindent
{\it Product gauge groups of the form $Sp(2k_1) \times SO(2k_2)
\times Sp(2k_3) \times \cdots \times SO(2k_n)$ with fundamental
matter}

The gauge group at the start  of the chain is arranged to be
$Sp(2k_1)$, while that at the end of the chain is chosen to be
$SO(2k_n)$, implying that $n+1$ is odd.
Demanding that there be no semi-infinite fourbranes, means that
there should be a solution $t\sim v^2$ as $v\to 0$, but no $t\to
\infty$ solution in this limit. Further imposing
symmetry under $v \to -v$ restricts the curve to the form
\eqn\prosixcu{
y^{n+1} +A_1(v^2) y^n + \cdots + v^2 A_{n+1} (v^2) =0~.
}
Substituting $z=P/y$, we find
\eqn\prosixz{
v^2 A_{n+1} z^{n+1} + A_n P z^n + \cdots + P^{n+1} = 0~.
}

The condition that there are no semi-infinite fourbranes leads to the
conditions that $A_\alpha P^{n+1-\alpha}$ is divisible by $A_{n+1}$
for
all $0\leq \alpha \leq n$. Following \supered\ we may then
solve for the $A_\alpha$
\eqn\polcoef{
A_\alpha = g_\alpha(v^2) \prod_{s=1}^{\alpha-1} J_s^{\alpha-s}~,
}
where $g_\alpha$ are polynomials and
\eqn\jfunc{
J_s = \prod_{a=i_{s-1} +1}^{i_s} (v^2 -e_a^2)~,
}
where the integers $i_\alpha$ are related to the number of sixbranes
between the $(\alpha\!-\!1)$-th and $\alpha$-th fivebrane by
$d_\alpha = i_\alpha-i_{\alpha-1}$.

Finally we must impose the additional constraint that arises for $Sp$
gauge groups  in the
case when we mod out by $v\to -v$ with $y$ and $z$ fixed,
that the curve \prosixcu\ have double
roots in $y$ at $v=0$. This fixes $n/2$ of the constants appearing
in the $g_\alpha$ for $\alpha=1 \cdots n$, in terms of the other
$n/2$. This gives us precisely the right number of parameters for the
curve \prosixcu\ to describe the Coulomb branch of the
$Sp(2k_1) \times SO(2k_2) \times Sp(2k_3) \times \cdots \times
SO(2k_n)$
gauge group with $(\bf 2k_\alpha, 2k_{\alpha+1})$ matter together with
$d_\alpha$ fundamentals in the $\alpha$ factor.

The case when $n+1$ is even may be described in a very similar way. Now
we have $Sp$ gauge groups both at the start and end of the chain.
The only change to \prosixcu\ is that $v^2 A_{n+1}$ is replaced by
$A_{n+1}$. The solution \polcoef\ is the same. The double root
condition
now fixes $(n+1)/2$ of the constant terms appearing in the
$g_\alpha$ for $\alpha=1,\cdots, n+1$, leaving us with the correct
number of parameters to describe the Coulomb branch of
$Sp(2k_1) \times SO(2k_2) \times Sp(2k_3) \times \cdots \times
Sp(2k_n)$ with $(\bf 2k_\alpha, 2k_{\alpha+1})$ matter together with
$d_\alpha$ fundamentals in the $\alpha$ factor.

\bigskip\noindent
{\it Product gauge groups of the form $SO(2k_1) \times Sp(2k_2)
\times SO(2k_3) \times \cdots SO(2k_n)$ with fundamental matter}

In this case we must have that $n+1$ is even.
To obtain the curve in this case we replace the
$y^{n+1}$ term
in \prosixcu\ by $v^2 y^{n+1}$. The solution for the $A_\alpha$
\polcoef\
is identical. The double root condition now fixes $(n-1)/2$ of the
constant
terms appearing in the
$g_\alpha$ for $\alpha=1,\cdots, n$. Thus we obtain the correct
number of parameters to describe the Coulomb branch of
$SO(2k_1) \times SO(2k_2) \times Sp(2k_3) \times \cdots \times
SO(2k_n)$ with $(\bf 2k_\alpha, 2k_{\alpha+1})$ matter and
$d_\alpha$ fundamentals in the $\alpha$ factor.

\bigskip\noindent
{\it Product gauge groups of the form $SO(2k_1+1) \times Sp(2k_2)
\times SO(2k_3+1) \times \cdots \times SO(2k_n+1)$ with fundamental matter}

When no semi-infinite fourbranes are present $n+1$ must be even
in order that the curve admit a $v\to -v$, $t\to -t$ symmetry.
The curve for this case takes the form
\eqn\prosixcu{
vy^{n+1} +A_1(v^2) y^n + \cdots + v A_{n+1} (v^2) =0~.
}
The same argument as above leads to the solution \polcoef\ for the
$A_\a$.
The curve contains the expected number of
parameters to describe the Coulomb branch of the theory.

\newsec{Elliptic Models}

In this section we will consider the generalization of the
elliptic models of \supered. Namely, we compactify the coordinate $x^6$
on a circle of radius $L$. At certain points on this circle
we place $n$ fivebranes, and suspended between the $(\alpha-1)$-th
and the $\alpha$-th fivebranes we have $2k_{\alpha}$ fourbranes. We
identify the $\a=0$ fivebrane with the one at $\a=n$.
An orientifold plane will extend in the $x^6$ direction. Let us
include also $2d_{\alpha}$ sixbranes localized between the
$\alpha-1$-th and $\alpha$-th fivebranes.

This configuration corresponds again to a product gauge group
of alternating orthogonal and symplectic groups. With a
compactified $x^6$ direction, consistency between flavor and
gauge groups forces the number of fivebranes $n$ to be even.
We can attempt to describe chains with even, or chains with odd orthogonal
gauge groups by setting the corresponding $k_{\alpha}$ to be
respectively integer or half-integer.

An important property of the elliptic models for unitary gauge
groups studied in \supered\ is that the product group
included an additional $U(1)$ factor not present in the non-compact
models, $G= U(1) \times \prod SU(k_{\alpha})$. The origin of the
extra $U(1)$ is that, for periodic configurations, equation \finite\
allows a global shift in the $v$ plane of the brane configuration.
In our case, even for elliptic models, this $U(1)$ is not present.
It is always eliminated by the orientifold projection since it
would correspond to moving the position of the orientifold plane
and this is not a dynamical degree of freedom of the theory.

We are interested in models with non-positive $\beta$-functions.
The sixbranes contribute a positive amount to the one-loop
coefficient of the $\beta$-functions
\eqn\ebetaM{b_{0,\alpha}=a_{\alpha} - a_{\alpha-1} + 2 d_{\alpha},}
with $a_{\alpha}$ given by \easP. If we want $b_{0,{\alpha}}\leq 0$
for each factor group $G_{\alpha}$, we have to set $d_{\alpha}=0$.
Therefore the matter content of these theories will consist of
half-hypermultiplets transforming in the $(\bf 2k_{\alpha},
2k_{\alpha+1})$ representations.

In the absence of sixbranes the condition $b_{0,{\alpha}} \leq 0$
reduces to $a_{\alpha-1} \geq a_{\alpha}$. This, together
with the periodicity of our configuration, implies
vanishing $\beta$-function coefficients and determines the
number of fourbranes to be
\eqn\ekE{k_{\alpha} = k - (1-(-1)^{\alpha})
{\omega_{\alpha} \over 2}.}
Let us fix $\omega_1=1$, the product group structure we get is
\eqn\eEP{ \cdots \times Sp(2k-2) \times SO(2k) \times Sp(2k-2)
\times \cdots.}
Only chains with even orthogonal groups can be derived from the
elliptic models.

In order to obtain the exact solution of these models, as before,
we lift our brane configuration to $M$-theory.
The compact $x^6$ and $x^{10}$ directions will
define now a Riemann surface of genus $1$, which we denote by $E$.
The elliptic curve $E$ can generically be described in terms
of two complex variables $x,y$ by
\eqn\eE{y^2 = 4 x^3 - g_2 x - g_3,}
with $g_2$, $g_3$ two complex numbers. The ambient space $Q$
describing the directions $x^4,x^5,x^6$ and $x^{10}$ is then
the product $E \times C$. In \supered\ non-trivial fibrations
of $C$ over $E$ were considered. They were associated with
configurations periodic in $x^6$ up to a shift in the $v$
coordinate. This shift translated into a bare mass parameter for
the gauge theory on the fourbranes.
Since it does not leave invariant the point $v=0$, the introduction
of a shift is not compatible with the orientifold projection.
Therefore we will only consider the direct product space
$Q=E \times C$. The absence of bare mass parameters for
orthogonal and symplectic gauge theories where the flavor group
has been gauged was already encountered
in section (3.5).

Following \supered, the Riemann surface solving the
elliptic models will be described by a $2k$-fold cover of the
elliptic curve $E$
\eqn\eCEP{F(x,y,v^2)= v^{2k} + f_1(x,y) v^{2k-2} + \cdots
+ f_k (x,y) =0,}
with each branch related to positions of the fourbranes in
the $v$ plane. Let us analyze what is necessary for 
\eCEP\ to represent the answer. For each factor $G_{\alpha}$ in
the product group, we denote by $p_{\alpha}$ the degree
in $v$ of the polynomial $F_{G_{\alpha}}(t,v^2)$ derived in
section 3. In the case of $SO(2k_1)$ groups
we have $p=2k_1$ and for $Sp(2k_2)$ we have instead
$p=2k_2 +2$.
Equation \eCEP\ can provide
the solution only when all the $p_{\alpha}$ in the cyclic
chain are equal, in particular $p_{\alpha}= 2k$. This is
precisely what we derive from the
non-positivity of the $\beta$-functions.

The presence of fivebranes is encoded in \eCEP\ in
terms of poles of the functions $f_i(x,y)$. More precisely,
each function $f_i$ will be chosen to have simple poles at $n$
points $p_1,\cdots,p_n$ representing the positions of the fivebranes.

To end this section we want to match the free parameters in \eCEP\
with parameters describing the moduli space of vacua of the gauge
theory. The positions of the fivebranes provide $n$ complex
parameters which correspond to the $n$ bare coupling constants
of our finite theory.
By the Riemann-Roch theorem, the space of meromorphic
functions with $n$ simple poles at $p_1,\cdots,p_n$ on a Riemann
surface of genus $1$ is $n$-dimensional.
Therefore \eCEP\ contains in addition $nk$ parameters.
The Coulomb branch of a gauge theory based on the product group
\eEP\ has dimension $nk - n/2$. Since we can not turn on bare
masses for the hypermultiplets, it seems that we get $n/2$
additional parameters without a gauge theory analogue.

This problem can be however solved by analyzing more carefully
the behavior of \eCEP\ at $v=0$. In sections (3.3) and (3.5)
we saw that the strong interaction between the orientifold
plane and the fivebranes gives rise to the condition that
at $v=0$ the fivebranes, pushed by the orientifold charge,
should meet pairwise. In terms of \eCEP\ this implies that
$f_k(x,y)$ should have $n/2$ double zeroes. A meromorphic function
with $n$ simple poles on a Riemann surface of genus $1$ will
have generically $n$ simple zeroes. We have thus to impose
that $f_k$ has $n/2$ double zeroes. This represents $n/2$ constraints
which eliminate $n/2$ parameters. Therefore the Riemann surface
\eCEP\ contains exactly the number of parameters we need to
represent the Casimirs of our gauge theory.

Finally let us comment on the Seiberg-Witten differential for the
elliptic models. As in section 2, this is obtained by considering the
area of a minimal volume membrane ending on the fivebrane. In this
case the area is given by
\eqn\susyellip{
V \sim \biggl| \int_D {dv ~dx \over y}\biggr| =
\biggl|\int_{\partial D}  {v ~dx \over y} \biggr|~,
}
and the Seiberg-Witten differential is $\lambda =v ~dx/y$.

{\bf Acknowledgments}

We would like to thank E. Gimon, J. Polchinski, J. Schwarz,
A. Strominger and N. Warner for discussions.
The research of K. L. is supported by the Fonds zur F\"orderung
der wissenschaftlichen Forschung, Erwin Schr\"odinger
Auslandsstipendium J01157-PHY and by DOE grant DOE-91ER40618.
E.L. is supported by a C.A.P.V. fellowship.
D.L. is supported in part by DOE grant DE-FG03-92-ER40701.

\listrefs
\end